\documentclass[preprint,12pt]{elsarticle}
\usepackage{graphicx}
\usepackage{lineno,hyperref}
\usepackage{natbib}
\usepackage{colortbl}
\usepackage{tabulary}
\usepackage{etoolbox}
\usepackage{multirow}
\usepackage{amsfonts}
\usepackage{amsmath}
\usepackage{xcolor}
\usepackage{lipsum}

\usepackage[numbers]{natbib}
\usepackage[shortlabels]{enumitem}
\newlist{abbrv}{itemize}{1}
\setlist[abbrv,1]{label=,labelwidth=0.4in,align=parleft,itemsep=-0.2\baselineskip,leftmargin=!}
\usepackage{subfigure}
\definecolor{gray95}{gray}{0.55}
\definecolor{gray25}{gray}{0.8}
\usepackage[ruled,vlined,lined,linesnumbered]{algorithm2e}
\usepackage[british]{babel}
\usepackage{caption}
\usepackage{mathtools}
\DeclarePairedDelimiter{\ceil}{\lceil}{\rceil}

\newenvironment{shrinkeq}[1]
{ \bgroup
  \addtolength\abovedisplayshortskip{#1}
  \addtolength\abovedisplayskip{#1}
  \addtolength\belowdisplayshortskip{#1}
  \addtolength\belowdisplayskip{#1}}
{\egroup\ignorespacesafterend}
\journal{Swarm and Evolutionary Computation}

\begin{document}

\begin{frontmatter}

\title{Utilizing Dependence among Variables in Evolutionary Algorithms for Mixed-Integer Programming: A Case Study on Multi-Objective Constrained Portfolio Optimization}                      




\address[mymainaddress]{Shanghai Key Laboratory of Multidimensional Information Processing, and the Department of Computer Science and Technology, East China Normal University, 3663 North Zhongshan Road, Shanghai 200062, China}
\address[mysecondaryaddress]{Electronics and Communication Sciences Unit, Indian Statistical Institute, 203, B. T. Road, Kolkata-700 108, India}

\author[mymainaddress]{Yi Chen}
\author[mymainaddress]{Aimin Zhou\corref{mycorrespondingauthor}}
\cortext[mycorrespondingauthor]{Corresponding author, amzhou@cs.ecnu.edu.cn. This work is supported by the Science and Technology Commission of Shanghai Municipality (No. 19511120600) and the National Nature Science Foundation of China (Nos. 61773296, 61673180).}

\author[mysecondaryaddress]{Swagatam Das}
\begin{abstract}
Several real-world applications could be modeled as Mixed-Integer Non-Linear Programming (MINLP) problems, and some prominent examples include portfolio optimization, remote sensing technology, and so on. Most of the models for these applications are non-convex and always involve some conflicting objectives. The mathematical and heuristic methods have their advantages in solving this category of problems. In this work, we turn to Multi-Objective Evolutionary Algorithms (MOEAs) for finding elegant solutions for such problems. In this framework, we investigate a multi-objective constrained portfolio optimization problem, which can be cast as a classical financial problem and can also be naturally modeled as an MINLP problem. Consequently, we point out one challenge, faced by a direct coding scheme for MOEAs, to this problem. It is that the dependence among variables, like the selection and weights for one same asset, will likely make the search difficult. We thus, propose a Compressed Coding Scheme (CCS), compressing the two dependent variables into one variable to utilize the dependence and thereby meeting this challenge. Subsequently, we carry out a detailed empirical study on two sets of instances. The first part consists of 5 instances from OR-Library, which is solvable for the general mathematical optimizer, like CPLEX, while the remaining 15 instances from NGINX are addressed only by MOEAs. The two benchmarks, involving the number of assets from 31 to 2235, consistently indicate that CCS is not only efficient but also robust for dealing with the constrained multi-objective portfolio optimization.
\end{abstract}
\end{frontmatter}

\section{Introduction}
A Mixed-Integer Non-Linear Programming (MINLP) model manifests itself in several real-world applications, ranging from the portfolio optimization~\cite{de2019distributionally}, path planning~\cite{yang2015path}, remote sensing technology~\cite{valicka2019mixed} to image classification~\cite{liu2013automatic}. For example, in the first two specific applications mentioned above, the selection (integer) and weight (continuous) of the assets, and the number (integer) and angle (continuous) of the rotations of an aircraft are mixed-integer variables that should be dealt with simultaneously. 

Without loss of generality, the majority of the discussion in this work is about the constrained portfolio optimization, because it could be naturally modeled as an MINLP problem and it is one of the well-known financial problems~\cite{ertenlice2018survey,kolm201460}. To be specific, a portfolio optimization problem considers an optimal weight of the limited fund in a series of risky assets, namely, securities, bonds, stocks, and derivatives. In practice, investors attempt to acquire the best-expected return at a given risk level or minimize the risk in an acceptable return range. In general, the expected return could be directly assessed by the profit. However, in terms of measuring the risk, there are different methods based on different assumptions of the markets, e.g., the Mean-Variance (MV)~\cite{van2020surprising}, the Value-at-Risk (VaR)~\cite{meng2020estimating}, and the Conditional Value-at-Risk (CVaR)~\cite{rockafellar2000optimization}. The MV model, playing a significant role in the progress of modern portfolio optimization, is studied as the basic model in this paper.

In the literature, some exact algorithms have been designed and used to solve the constrained portfolio optimization problems~\cite{2009BertsimasACCQO}. Nonetheless, challenges emerge when the problems involve non-convex quadratic models or the total number of combinations of assets are large~\cite{estrada2008mean,shaw2008lagrangian}. On the other hand, population-based Evolutionary Algorithms (EAs) can also tackle these problems~\cite{2000TJChangHFCCPO, kalayci2020efficient}. Generally, the accuracy of the EAs will be worse than that of the mathematical methods, but EAs do possess a better versatility and can be applied to both convex and non-convex problems. We turn to EAs in this paper.

There are many variations of portfolio problems based on the MV model. Concerning the objective function, they could be roughly divided into three categories as follows:
\begin{itemize}
\item[(i)] \emph{weighted formulation:} it combines the two objectives (risk and return) by using a weighting parameter and regards it as the final objective. To the best of our knowledge, it plays a dominant role in the single objective MV model for portfolio optimization~\cite{2000TJChangHFCCPO, 2011GasperoHybrid, jahan2010local}.
\item[(ii)] \emph{transforming objective functions:} it transforms one of the objectives into an equality or inequality constraint and considers the other as the final objective~\cite{2002SchaerfLocal}, or integrates the considered objectives into one with some criteria~\cite{pouya2016solving}.
\item[(iii)] \emph{multi-objective models:} generally, they regard the risk and return as two main aspects, and aim to find a set of trade-off solutions~\cite{branke2009portfolio, 2014KhinMODEwawl, mishra2014comparative}. Furthermore, they can involve more than two objectives when considering more issues~\cite{saborido2016evolutionary}.
\end{itemize}

Following the above work, this paper considers an extended MV model that incorporates four real-world constraints~\cite{2014KhinMODEwawl}: (i) \emph{cardinality constraint:} it restricts the number of assets in the portfolio result. (ii) \emph{floor and ceiling constraint:} it determines the minimal and maximal quantities of every asset. (iii) \emph{pre-assignment constraint:} it considers the preference of investors. (iv) \emph{round lot constraint:} it demands the holding quantities of assets should be multiple of the minimal round lot. This constrained portfolio optimization has two layers of optimization~\cite{li2020alternative}. In the first level, it aims to find the best selection (combination) of available assets, which contains 0-1 integer variables, i.e., an asset is chosen or not. In the second level, it aims to find an optimal weight of a finite fund, which contains continuous variables, i.e., the proportion of the fund assigned to each asset. Hence, the extended MV model can be converted into an MINLP problem and it has been proved to be NP-hard~\cite{gao2013optimal, 2013Newmantutorial}. How to deal with 0-1 integer and continuous variables has become a key issue when using an EA. It is natural to adopt a binary vector to specify the selection of the assets and a real-valued vector to indicate the investment proportions. This kind of direct representation (coding scheme) is the most popular strategy with EAs~\cite{canakgoz2009mixed,lwin2017mean}. However, the direct coding scheme also leads to challenges to algorithm design, and more specifically, (1) it is hard to reuse existing search operators in EAs since they are usually designed either for continuous or discrete variables, and (2) the dependence, among the selection and weight, of the variables probably slow down the search.

Facing these challenges, in this paper, we propose a Compressed Coding Scheme (CCS), with which merely one real-valued vector is employed to represent the selection and weight simultaneously. In this way, not only the reusing of the existing search operators is simplified, but also the dependence can be utilized for a more effective search. In fact, some prior works have already mentioned the use of a real-valued vector to represent both discrete and continuous variables simultaneously~\cite{chen2017evolutionary, li2013multi}. However, to the best of our knowledge, this article is the first to discuss this coding scheme in-depth, and also, it points out the above two advantages for the first time. Last but not least, some tailored search operators are proposed to enhance the performance of this coding scheme for this constrained portfolio optimization problem.

Further, it has been pointed out as a matter of fact that the objectives, viz. the expected return and risk of portfolios always conflict with each other~\cite{ponsich2013survey}. In such problems, the target is to find a set of solutions that could represent the best possible trade-off among the objectives, instead of identifying an optimal solution. Hence, CCS is integrated into three existing state-of-the-art Multi-objective Evolutionary Algorithms (MOEAs), i.e., Decomposition based Multi-objective Evolutionary Algorithm (MOEA/D)~\cite{2007QingfuMOEAD}, Non-dominated Sorting Genetic Algorithm (NSGA-\uppercase\expandafter{\romannumeral2})~\cite{2002DebNSGA2}, and $\varphi$ Metric Selection Evolutionary Multi-objective Algorithm (SMS-EMOA)~\cite{2006NicolaSMS}. In a series of instances, we have conducted several simulation studies, and our experimental results suggest that MOEAs with CCS exhibit higher efficiency and robustness in searching for optimal solutions. These solutions are superior for their better diversities and shorter distances to the Pareto Front (PF). 

The structure of this article is presented as follows. Section~\ref{section_2} introduces the formulation of the constrained multi-objective portfolio optimization problem. In Section~\ref{section_3}, the direct and compressed coding schemes are presented. In Section~\ref{section_4}, some reproduction operators and a repairing method are introduced. Then, a complete algorithm framework, including CCS and a multi-objective selection method, is presented. Thereafter, some simulation experiments are presented in Section~\ref{section_5}. Finally, Section~\ref{section_6} outlines the conclusions of this paper and presents some future research directions.

\section{Mathematical Model}
\label{section_2}

Before the definition of the constrained portfolio optimization model, we give the following notations.
\begin{abbrv}
\item[$N$]                    the number of available assets
\item[$K$]                    the number of assets in a portfolio, i.e., the cardinality
\item[$L$]                    the number of assets in the pre-assignment set
\item[$w_i$]                 the proportion of capital invested in the $i$-th asset
\item[$\rho_{ij}$]       the correlation coefficient of the $i$-th and $j$-th assets
\item[$\sigma_{i}$]     the standard deviation of $i$-th asset
\item[$\sigma_{ij}$]   the covariance of $i$-th and $j$-th assets
\item[$\mu_i$]              the expected return of the $i$-th asset
\item[$\upsilon_i$]       the minimum trading lot of the $i$-th asset
\item[$\epsilon_i$]       the lower limit on the investment of the $i$-th asset
\item[$\delta_i$]          the upper limit on the investment of the $i$-th asset
\item[$y_i$]                  the multiplier of the minimum trading lot in the $i$-th asset
\end{abbrv}
\begin{shrinkeq}{-2ex}
\begin{flalign*}
&\sigma_{ij} = \rho_{ij}\sigma_{i}\sigma_{j}
&
\end{flalign*}
\end{shrinkeq}
\begin{shrinkeq}{-2ex}
\begin{flalign*}
&s_i=
\begin{cases}
1, &\textbf{if} \ $the $i$-th $(i=1,...,N)$ asset is chosen,$ \cr 0, &$otherwise,$
\end{cases}
&
\end{flalign*}

\end{shrinkeq}
\begin{shrinkeq}{-1ex}
\begin{flalign*}
&z_i=
\begin{cases}
1, &\textbf{if} \ $the $i$-th asset is in the pre-assigned set,$ \cr 0, &$otherwise.$
\end{cases}
&
\end{flalign*}
\end{shrinkeq}

This paper considers the following bi-objective model~\cite{1996BienstockCSOFMQPP}, which includes maximizing the return and minimizing the risk simultaneously. Meanwhile, it meets the four practical constraints~\cite{2014KhinMODEwawl} mentioned above, namely, cardinality, quantity, pre-assignment, and round lot constraints.
{
\setlength{\abovedisplayskip}{3pt}
\setlength{\belowdisplayskip}{3pt}

\begin{flalign}
\label{risk}
&min\quad f_1=\sum_{i=1}^{N}\sum_{j=1}^{N}w_{i}w_{j}\sigma _{ij},&
\end{flalign}
\begin{flalign}
\label{return}
&max\quad f_2=\sum_{i=1}^{N}w_{i}\mu _{i},&
\end{flalign}
\begin{flalign}
\label{sum_to_one}
&subject \quad to \quad \sum_{i=1}^{N}w_{i}= 1,\qquad 0\leq w_{i}\leq 1,&
\end{flalign}
\begin{flalign}
\label{Cardinality Constraint}
&\sum_{i=1}^{N}s_{i}=K,&
\end{flalign}
\begin{flalign}
\label{floor_ceiling}
&\epsilon _{i}s_{i}\leq w_{i}\leq \delta _{i}s_{i},\qquad i=1,...,N,&
\end{flalign}
\begin{flalign}
\label{Pre-assignment}
&s_{i}\geq z_{i},\qquad i=1,...,N,&
\end{flalign}
\begin{flalign}
\label{Round Lot}
&w_{i}=y_{i}\tau,\qquad i=1,...,N,\quad y_{i}\in \mathbb{Z}_{+},&
\end{flalign}
\begin{flalign}
\label{discrete_binary}
&s_{i}\in \{0,1\},\qquad i=1,...,N,&
\end{flalign}
}where Eqs.~(\ref{risk}) and (\ref{return}) are two respective objectives, minimizing the risk and maximizing the return, in the portfolio optimization that conflict with each other. Eq.~(\ref{sum_to_one}) requires that all the capital should be invested in a valid portfolio. Eq.~(\ref{Cardinality Constraint}) is the \emph{cardinality constraint} (i.e., $K$ assets are selected), and Eq.~(\ref{floor_ceiling}) is the \emph{floor and ceiling constraint}, which restricts the investment proportion being allocated in the $i$-th asset should lie in [$\epsilon _{i} ,\, \delta _{i}$]. In addition, Eq.~(\ref{Pre-assignment}) represents that the $i$-th asset must be included in a portfolio ($z_i=1$), when it is of interest for the investor. It is a \emph{pre-assignment constraint}. Thereafter, Eq.~(\ref{Round Lot}) defines the \emph{round lot constraint}. The round lot size $\tau$ is set to be a same constant, with which 1 is divisible, since it will be really hard to handle a flexible one. In that case, it will be beyond the scope of the discussion about coding schemes in this work. Finally, Eq.~(\ref{discrete_binary}), which is the~\emph{discrete constraint}, implies that both the $s_i$ and $z_{i}$ must be binary

\section{Coding Schemes}
\label{section_3}

The decision variables in the model presented in the last section include $s_i\in \{0,1\}$ and $w_i\in [0,1]$ for $i=1,..., N$. When using EAs to deal with such mixed variables, the solution representation or coding scheme becomes vital. This section firstly introduces a popular Direct Coding Scheme (DCS)~\cite{2014KhinMODEwawl,1994BeanGenetic}. We point out that there will be an explicit dependence among variables. Then, a new coding scheme, called Compressed Coding Scheme (CCS), is introduced for EAs to utilize the dependence via grouping dependent variables.

\subsection{Direct Coding Scheme (DCS)~\cite{2014KhinMODEwawl}}

\begin{figure}[htbp]
\centering 
 \graphicspath{{figs/}}
 \includegraphics[width=0.4\columnwidth]{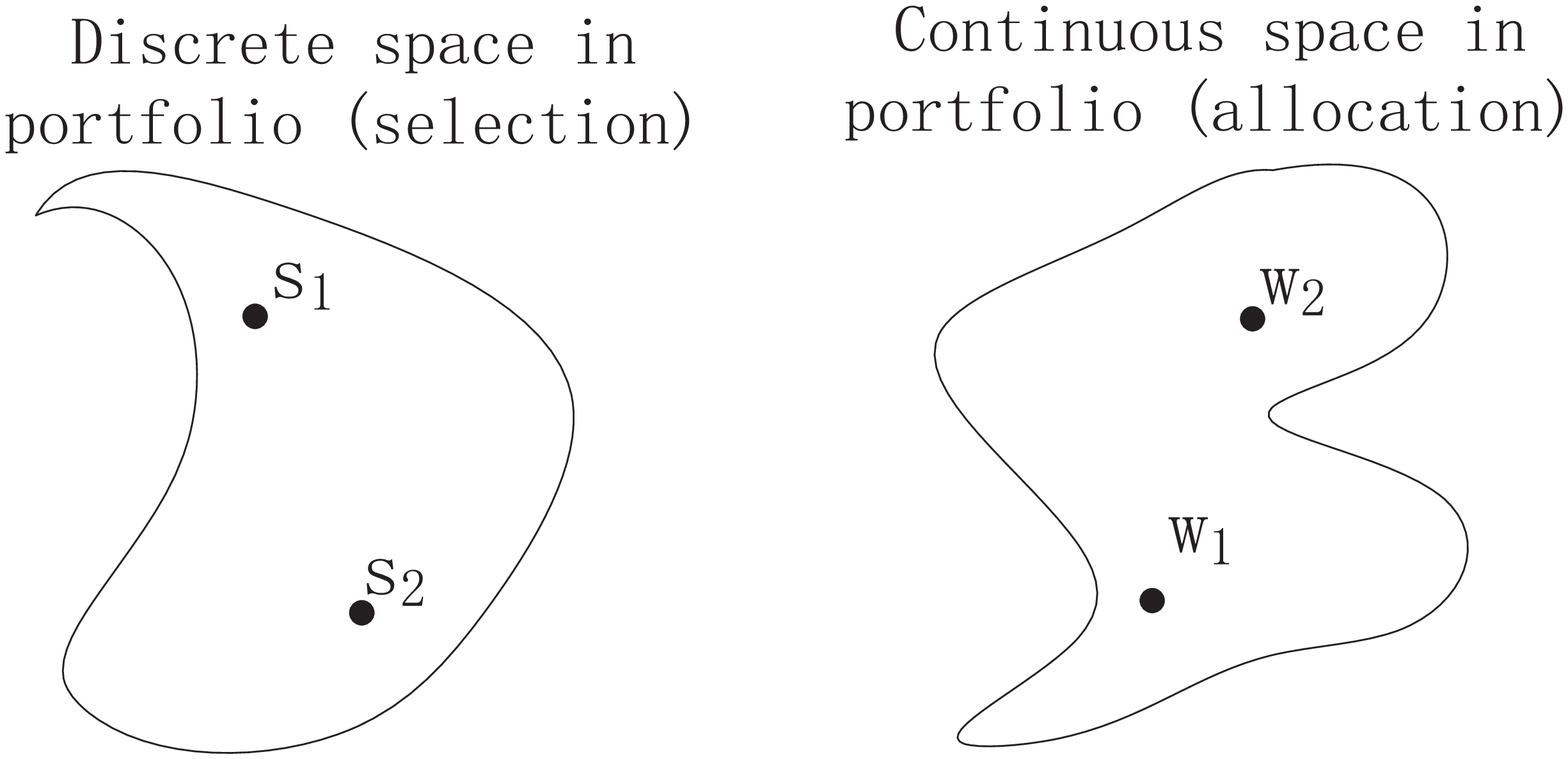}
  \caption{An illustration of DCS in portfolio optimization. Principally, there are two search spaces in a constrained portfolio optimization problem (discrete space and continuous space respectively). With DCS, an EA finds the selection and weight of the assets directly in the actual solution spaces. In the discrete space, $s_1$ or $s_2$ is a point that presents the combinations of assets for selection, such as $\{2,4\}$ with binary vector $\{0,1,0,1,0\}$. In the continuous space, $w_1$ or $w_2$ is a point that presents the weights of allocation, like real-valued vector$\{0.23,0.77\}$.} 
  \label{Direct_Mapping}
\end{figure}

In DCS, a solution is represented by a vector as follows:
\begin{flalign}
\label{chromosome_double_formulation}
&\textbf{c}=\Big( c_1,c_2,...,c_N,...,c_{2N}\Big),& 
\end{flalign}
where $c_i\in \{0,1\}$, $i=1,...,N$ and $c_i\in[0,1]$, $i= N+1,...,2N$. It is clear that a solution includes two parts, i.e., the selection vector $(c_1,\cdots, c_N)$ and the weight vector $(c_{1+N},\cdots, c_{2N})$ respectively. Fig.~\ref{Direct_Mapping} illustrates this coding scheme.

\begin{figure}[htbp]
\centering
\graphicspath{{figs/}}
  \label{decoding:subfig:RK} 
  \includegraphics[width=0.6\columnwidth]{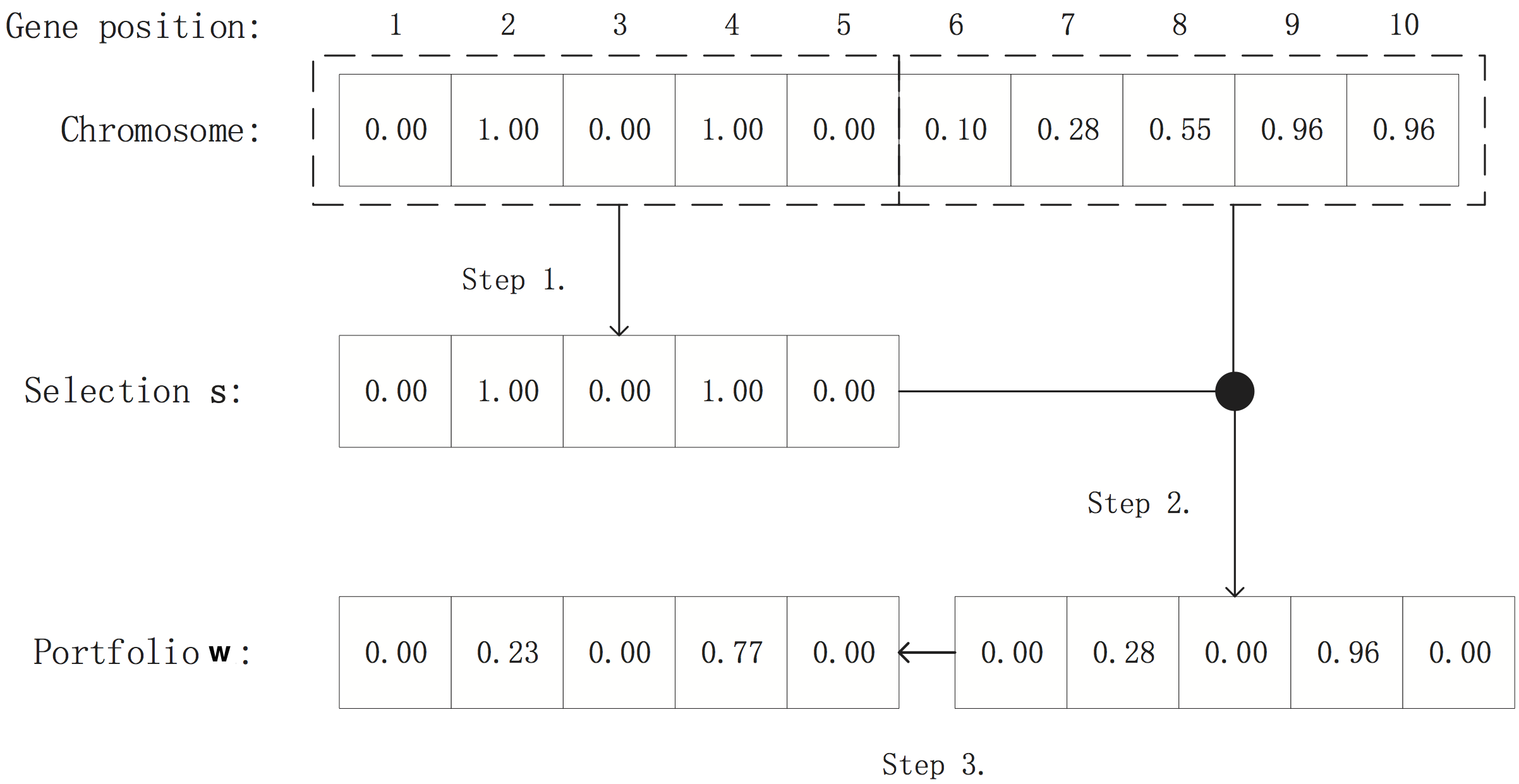}
  \caption{An example of decoding with DCS. The length of this solution is twice of the amount of available assets. Suppose $N=5$ and $K=2$ here, the decoding process can be divided into three steps. Step 1: The binary vector $\{0,1,0,1,0\}$ indicates the selection of assets is $s_2=1$ and $s_4=1$. Step 2: According to the selection \textbf{s}, the weight vector in the solutions is \{0.00, 0.28, 0.00, 0.96, 0.00\}. Step 3: Finally, the portfolio $\textbf{w}$ is normalized as \{0.00, 0.23, 0.00, 0.77, 0.00\}.} 
 \label{direct_decoding} 
\end{figure}

As for the decoding process, a solution is converted into the selection of assets $s_i$ as in Eq.~(\ref{direct_s}) and the weight of the assets as in Eq.~(\ref{decoding_RK}). An example of decoding is presented in Fig.~\ref{direct_decoding}. 
\begin{flalign}
\label{direct_s}
&s_i=
\begin{cases}
1, &\textbf{if} \ c_i=1 \cr 0, &$otherwise$
\end{cases}
,\qquad i = 1,2,...,N&
\end{flalign}
\begin{flalign}
\label{decoding_RK}
&w_i=\frac{s_ic_{i+N}}{\sum_{j=1}^{N}{s_jc_{j+N}}}, \qquad i = 1,2,...,N.&
\end{flalign}

\begin{figure}[htbp]
\graphicspath{{figs/}}
\centering
\tiny
\begin{minipage}{0.45\linewidth}
\centerline{\includegraphics[width=1\columnwidth]{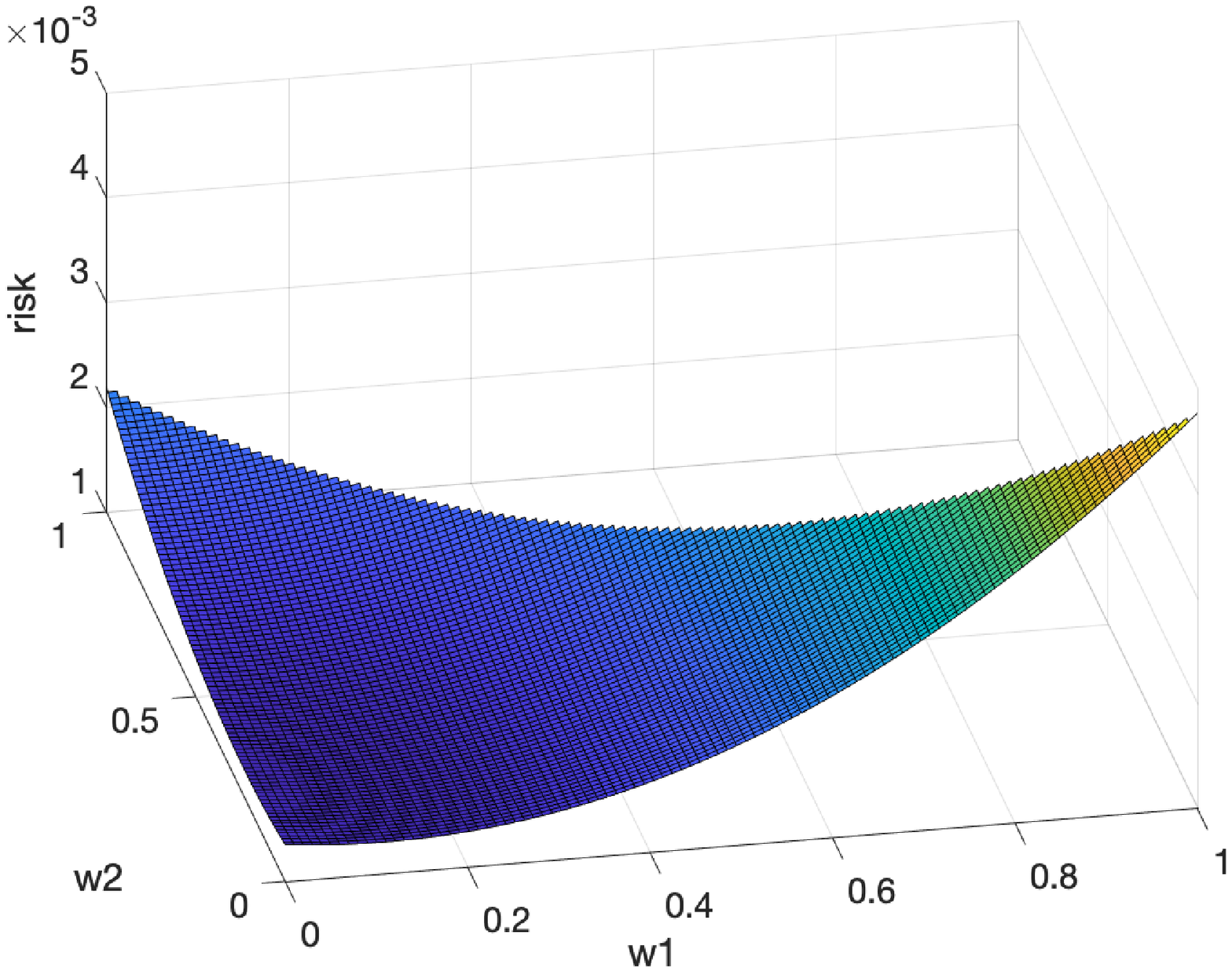}}
\centerline{(a) 3-D risk for portfolio1}
\end{minipage}
\begin{minipage}{0.45\linewidth}
\centerline{\includegraphics[width=1\columnwidth]{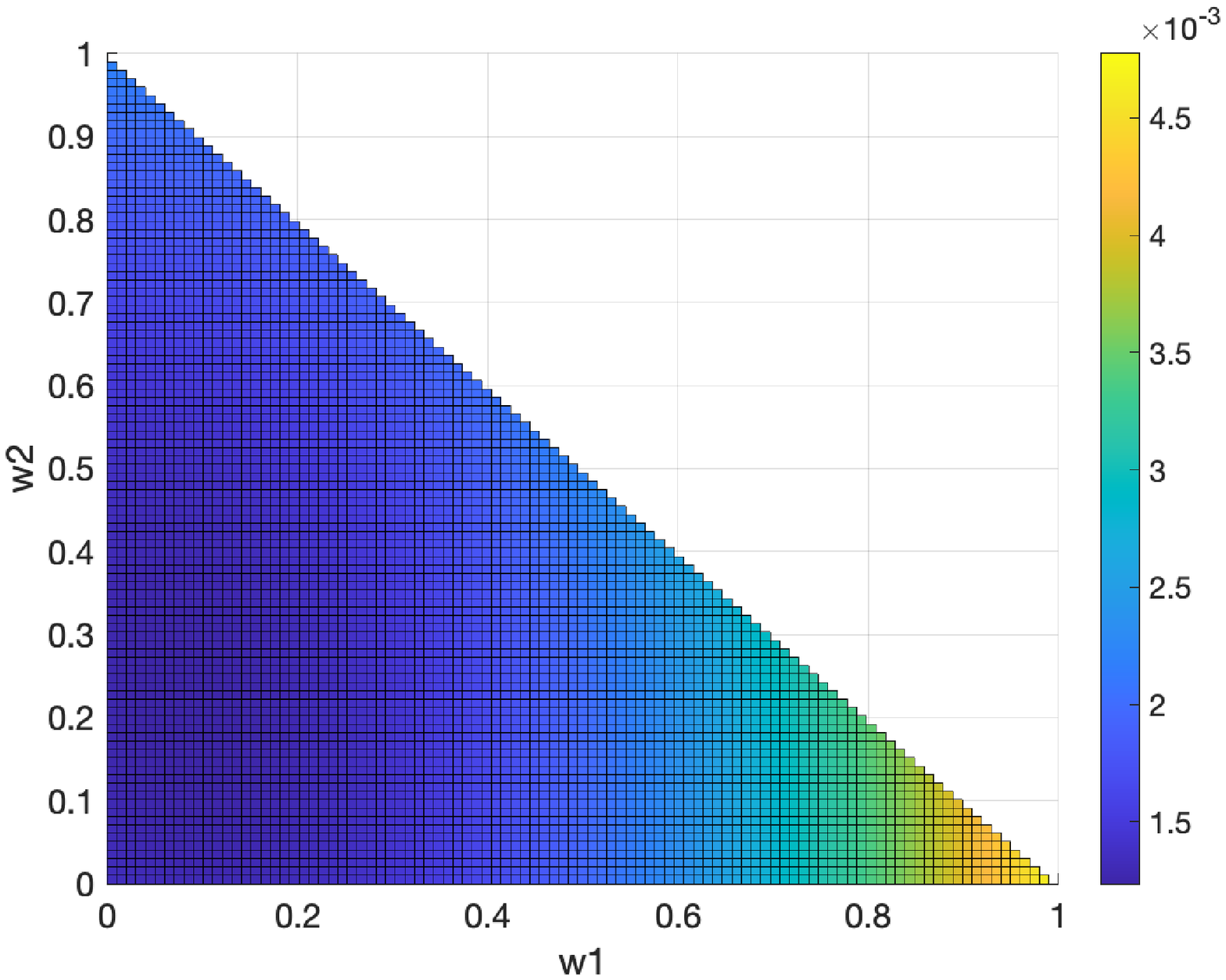}}
\centerline{(b) Heatmap risk for portfolio1}
\end{minipage}
\begin{minipage}{0.45\linewidth}
\centerline{\includegraphics[width=1\columnwidth]{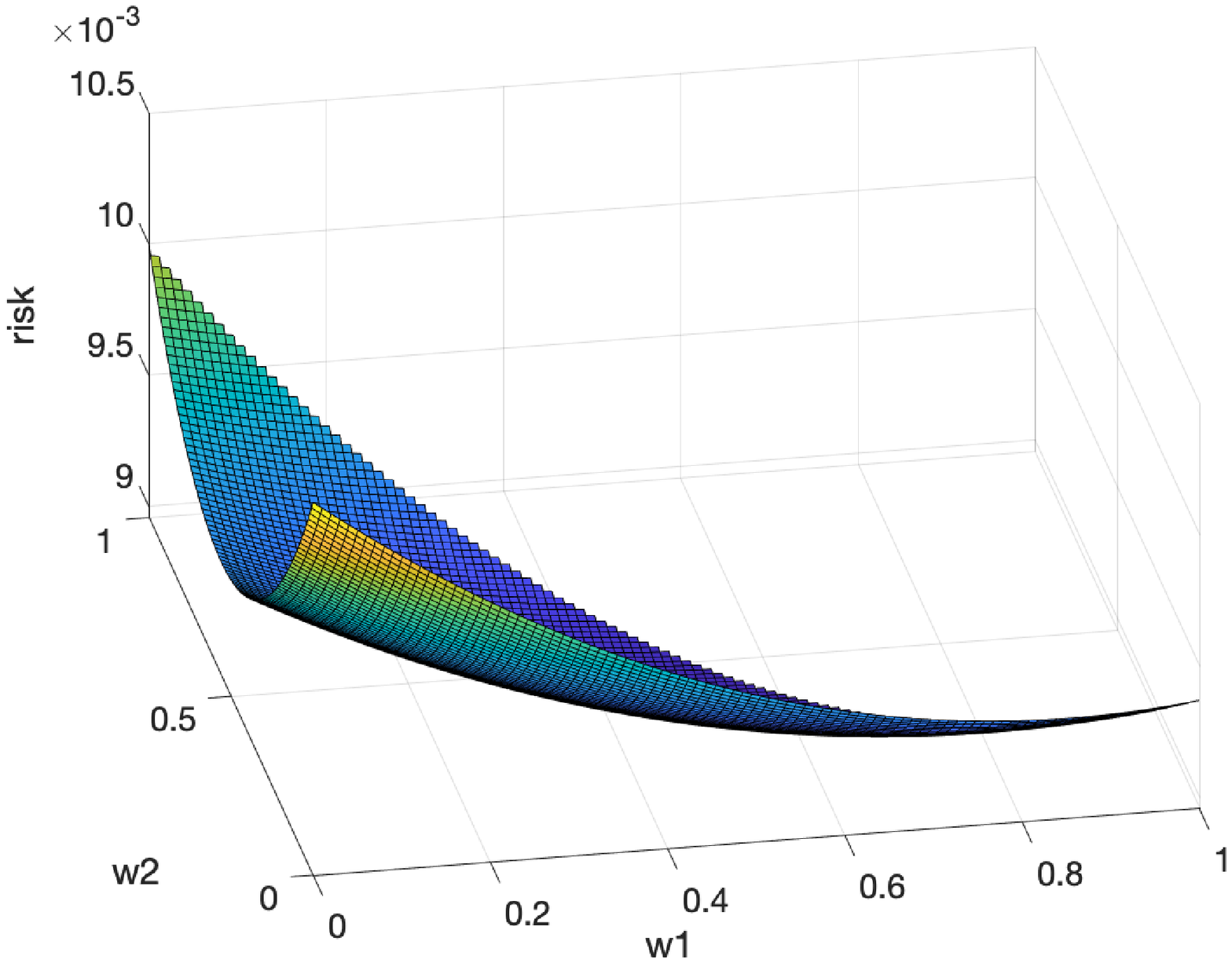}}
\centerline{(c) 3-D risk for portfolio2}
\end{minipage}
\begin{minipage}{0.45\linewidth}
\centerline{\includegraphics[width=1\columnwidth]{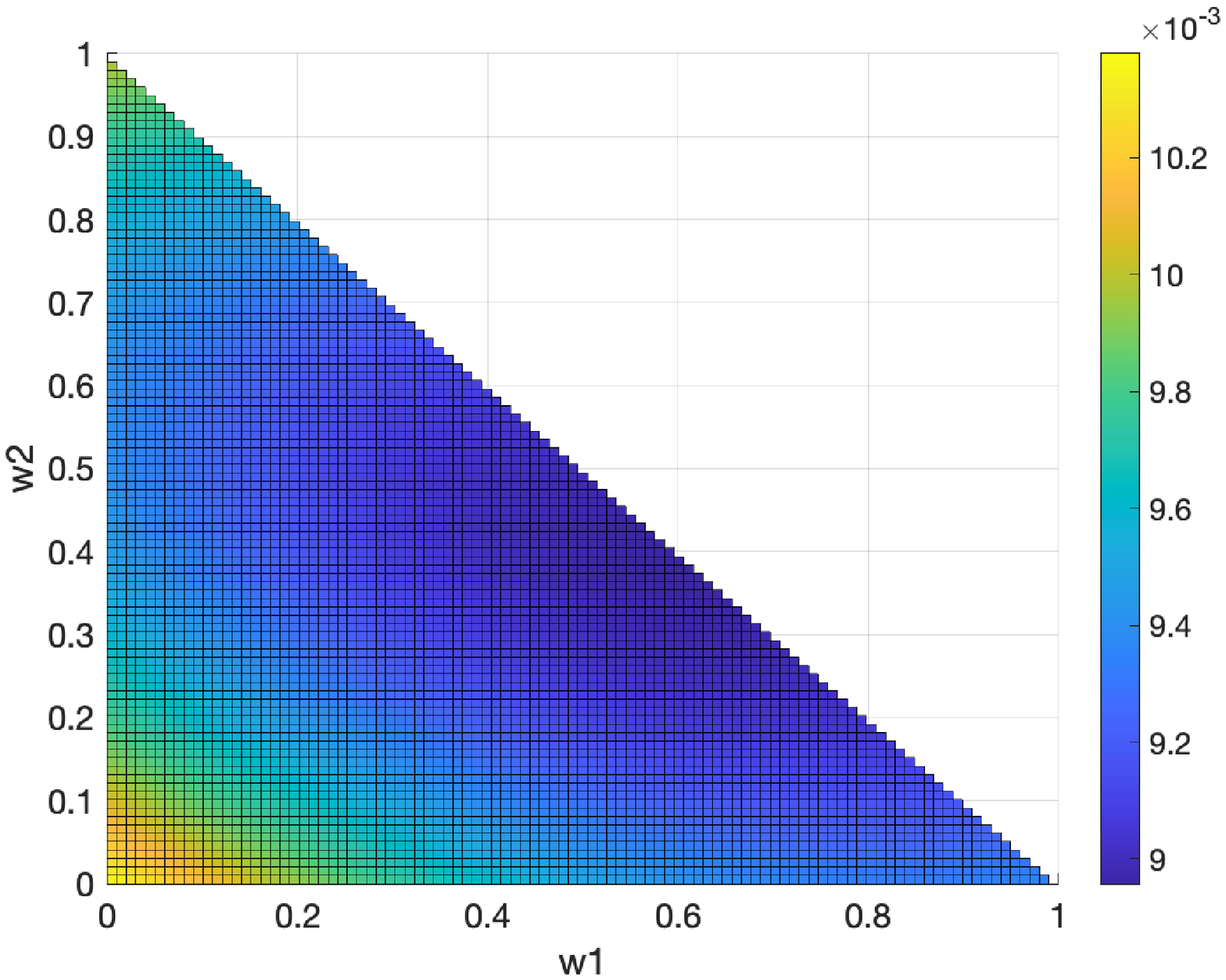}}
\centerline{(d) Heatmap risk for portfolio2}
\end{minipage}
\begin{minipage}{0.45\linewidth}
\end{minipage}
 \caption{The risk landscapes of two portfolios, portfolio1 and portfolio2. Each portfolio consists 3 different assets, of which the weights are $w1$, $w2$ and $w3$ respectively, where $w3=1-w1-w2$.} 
 \label{dependence}
\end{figure}

The dependence among variables in EA is not trivial. Some works~\cite{omidvar2017dg2, salomon1996re, yang2008large} have demonstrated that the dependence probably makes it difficult for EAs to search for the optimal solutions. This dependence will also occur on DCS since there is a non-linear relationship among $w_i$, $c_i$ (selection) and $c_{N+i}$ (weight), as shown in Eqs.~\ref{direct_s} and~\ref{decoding_RK}. Further, Fig.~\ref{dependence} shows the risk landscapes for two portfolios with different assets. Apparently, different combinations of assets make up different risk landscapes. For example, lowest risk of the portfolio1 at $w1=w2=0$ and $w3=1$. On the contrary, lowest risk of the portfolio2 at $w1=w2=0.5$ and $w3=0$. It is similar to the return landscapes of portfolios. These denote there is dependence among the selections and weights since good selections and weights are paired. This is hardly considered for designing EAs for portfolio problems. Contrarily, in the next subsection, we propose CCS for using only one real-valued vector to represent both selection and weight simultaneously. It is expected that EAs can utilize the dependence among variables when they are artificially grouped together. Note that the description about DCS in this subsection is a canonical one, involving one binary vector and one real-valued vector. Concerning the experimental study latter, a DCS, which utilizes two real-valued vectors, is also employed for reducing the gap between DCS and CCS. Therefore, we can make a fair comparison between them.

\subsection{Compressed Coding Scheme (CCS)}
\label{CCS_section}

\begin{figure}[htbp]
 \centering 
 \graphicspath{{figs/}}
\includegraphics[width=0.3\columnwidth]{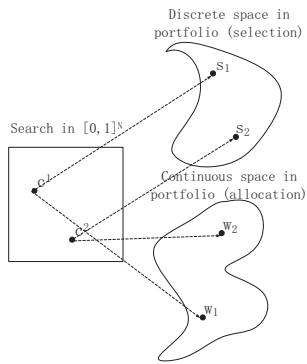}
  \caption{The illustration of CCS in portfolio optimization. Principally, there are two search spaces in a constrained portfolio optimization problem (discrete space and continuous space respectively). In the discrete space, $s_1$ or $s_2$ is a point that presents the combinations of assets for selection, such as $\{2,4\}$. In the continuous space, $w_1$ or $w_2$ is a point that presents the weights of allocation, like $\{0.23,0.77\}$.} 
  \label{CCS_Mapping}
\end{figure}

This section introduces a new coding scheme, CCS, for the EAs on the portfolio problem. The basic idea is to use a real-valued vector with length $N$ to represent a solution, which is defined in Eq.~(\ref{chromosome_CCS}).
\begin{flalign}
\label{chromosome_CCS}
&\textbf{c}=\Big( c_1,c_2,...,c_N\Big),& 
\end{flalign}
where $c_i\in[0,1]$, and $ i = 1,2,...,N$. The length of the solution is as the same as the number of assets. Fig.~\ref{CCS_Mapping} shows that CCS represents the selection and weight based on one string of real numbers in $[0,1]^N$ for a multi-mapping, where one vector $\textbf{c}$ is able to represent both selection and weight tasks simultaneously.

\begin{figure}[htbp]
\graphicspath{{figs/}}
  \label{decoding:subfig:CCS} 
  \centering
  \includegraphics[width=0.6\columnwidth]{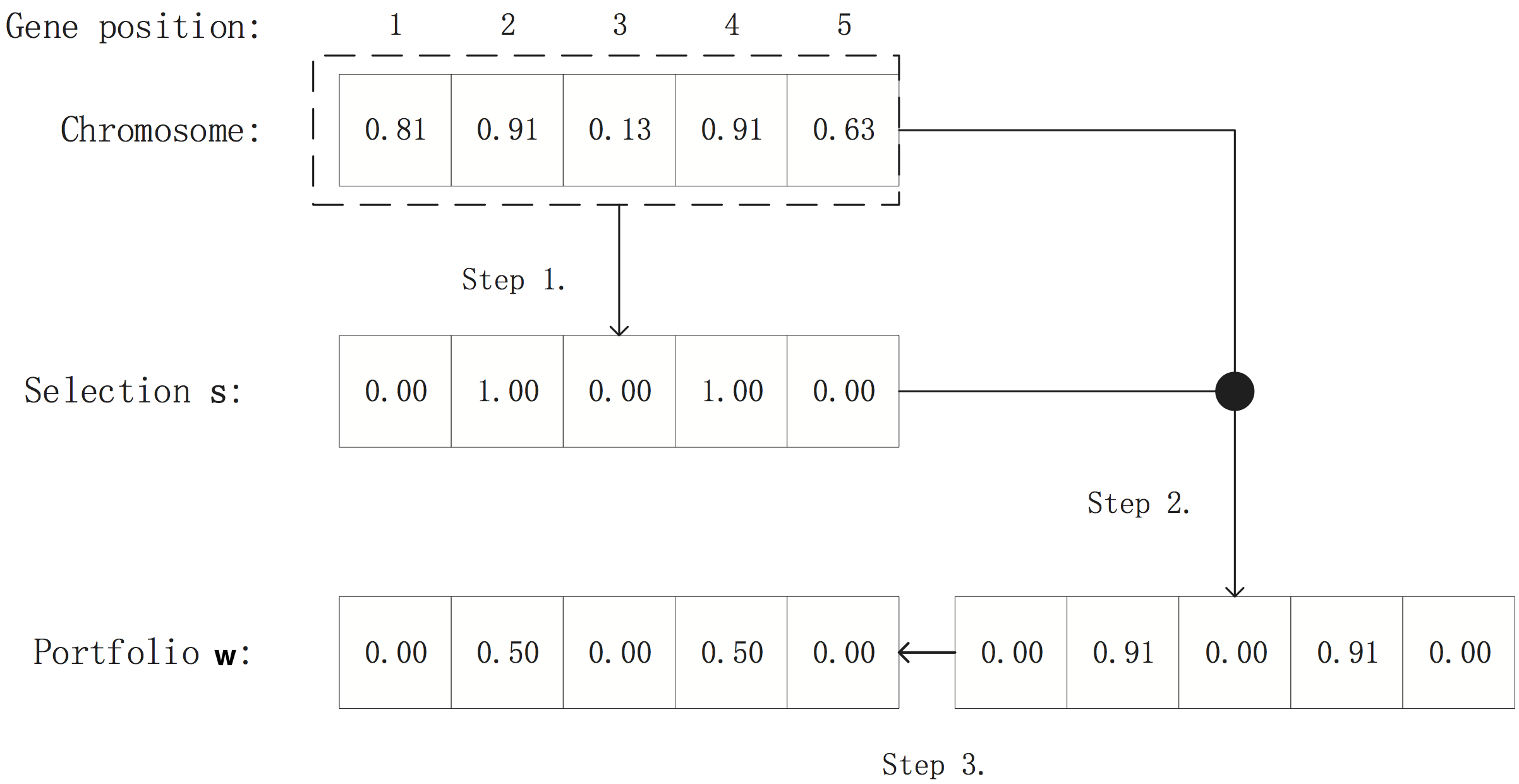}
  \caption{The example of decoding with CCS. In contrary to the decoding mentioned above, the length of this solution is as the same as the number of available assets and this solution is actually used twice. Suppose $N=5$ and $K=2$ here, the decoding process can be divided into three steps. Step 1: The values of $c_2$ and $c_4$ are higher than other genes, so the selection of assets is represented as $s_2=1$ and $s_4=1$. Step 2: Since the solution is utilized twice, the strings, which will interact with each other, are \{0.00, 1.00, 0.00, 1.00, 0.00\} and \{0.81, 0.91, 0.13, 0.91, 0.63\} respectively. So far, the weight before normalization is \{0.00, 0.91, 0.00, 0.91, 0.00\}. Step 3: Finally, the portfolio $\textbf{w}$ is normalized as \{0.00, 0.50, 0.00, 0.50, 0.00\}.} 
 \label{CCS_decoding} 
\end{figure}
The elements of a vector are utilized to represent not only the selection but also the weight. More specifically, the $L$ pre-assigned assets are selected. Meanwhile, among the assets which are not pre-assigned, the $K-L$ assets with highest values are selected. Then the $K$ elements of the vector are applied \emph{again} to represent the weight. The decoding process works as follows. Firstly, a solution is converted into the selection of assets ${s_1,...,s_N}$ where $s_i=1$ if the $i$-th asset is pre-selected or $c_i$ corresponds to the $K-L$ highest values except for the pre-selected assets. Then the weight of the assets is given in Eq.~(\ref{decoding_ccs}). An example of decoding is presented in Fig.~\ref{CCS_decoding}. 
\begin{flalign}
\label{decoding_ccs}
&w_i=\frac{s_ic_i}{\sum_{j=1}^{N}{s_jc_j}}, \qquad i = 1,2,...,N.&
\end{flalign}
\section{CCS based Algorithm Framework}
\label{section_4}

This section introduces how to deal with the constrained portfolio optimization with CCS based EAs. Firstly, three search operators are presented. Then, a repairing method is proposed to make all solutions feasible. Finally, a complete algorithm framework is given.

\subsection{Search Operators}
\label{search_operators}
By using CCS, all solutions are represented as real-valued vectors. Therefore, the search operators for continuous variables can be directly employed here. As suggested in~\cite{2014KhinMODEwawl}, this work firstly adopts a differential evolution (DE) strategy~\cite{1996RainerDE} and a polynomial mutation~\cite{deb2014analysing}, which never use prior knowledge.
\begin{abbrv}
\item[\textbf{O1}]         $c_i=c3_i+F\times (c1_i-c2_i)\\c'_i={\rm polymt}(c_i,\eta_m,p_m)$
\end{abbrv}
where $c1$, $c2$ and $c3$ are three solutions randomly selected from the population, $F$ is a scaling factor in DE, ${\rm polymt}(c_i,\eta_m,p_m)$ is a polynomial mutation function on $c_i$ with an index parameter $\eta_m$ and a mutation probability $p_m$, and $c'$ is the new solution.

Concerning the properties of the problems and CCS, we also propose two new search operators. Firstly, it is observed that the gene values will reserve lots of large but similar values when the number of assets is large since using the rank method for selection. The following search operator utilizes this heuristic information.
\begin{abbrv}
\item[\textbf{O2}]         $c'_i={c_i}^{r(1,2)}$
\end{abbrv}
where $r(1,2)$ is a random number in $[1,2]$. This operator only changes the investment proportion, but keeps the combination of assets, because the rank of values for genes is not changed. Further, this operator may also increase the variance of values in the solution vector, which can let the smallest weight close to 0~\cite{chen2018evolutionary}.

Secondly, a tailored operator that utilizes the known information of the portfolio optimization problem is proposed.
\begin{abbrv}
\item[\textbf{O3}] Swap the values of $c_i$ and $c_j$: $c_i\rightleftharpoons c_j$, where the asset $i$ is randomly chosen from the selected assets and the asset $j$ is chosen by randomly using one of the following strategies
\end{abbrv}
\begin{itemize}
\item Randomly choose another asset from the selected assets
\item Choose an asset which has the lowest correlation coefficient ($\rho_{jj}$) value
\item Choose an asset which has the highest return ($\mu_j$) value
\item Choose an asset which has the least correlation 
\\with those $K-1$ assets already chosen ($\mathop{\arg\min}_{j} \sum_{k\in {\rm selection}/\{i\}}{\rho_{kj}}$)
\end{itemize}

\subsection{Constraints Handling Method}
\label{constraint_handling}
New candidate solutions are repaired by slightly altering a method from~\cite{2014KhinMODEwawl} if the quantity and round lot constraints are violated. The procedure works as follows.

1. All weights for selected assets that are smaller than the value of $\ceil[\big]{\epsilon_{i} /\upsilon_{i}}\upsilon_{i}$ are adjusted by setting $w_{i}$:=$\ceil[\big]{\epsilon_{i} /\upsilon_{i}}\upsilon_{i}$ (the floor and round lot constraints). 

2. The weights are rounded down as $w_{i}$=$w_{i}$-($w_{i}$ mod $\upsilon_{i}$).

3. Alterations for making $\sum_i^K{w_i}= 1$ (the sum to one constraint).
\begin{itemize} 
\item If $\sum_i^K{w_i}>1$, decreasing the maximum multiples $\tau$ for the largest weight while meeting the floor and ceil constraints.

\item If $\sum_i^K{w_i}<1$, increasing the maximum multiples $\tau$ for the smallest weight while meeting the floor and ceil constraints.
\end{itemize}

4. If $\sum_i^K{w_i}\neq1$ then go to step 3, else the repairing process ends.

\subsection{Algorithm Framework}

This subsection introduces an algorithm framework for dealing with the constrained portfolio optimization problems, including CCS, the search operators, and the constraint handling method. The detailed pseudo-code of the algorithm framework is presented in Algorithm~\ref{ALG1}. 

\begin{algorithm}[h]
\footnotesize
\SetAlgoLined
\LinesNumbered

\tcp{\textcolor{blue}{initialization}}

Sample the initial population $P$ randomly in $[0,1]^N$.

\ForEach{$\textbf{c} \in P$}
{
\tcp{\textcolor{blue}{decode and repair}} 

Decode $\textbf{c}$ and repair it for $\textbf{w}$ if the constraints are violated.

\tcp{\textcolor{blue}{evaluation}}

Evaluate $\textbf{c}$ with $\textbf{w}$.
}

\While {\emph{stopping criteria is not met}}
{

Set $P' = \emptyset$.

\ForEach{$\textbf{c} \in P$ }
{
\tcp{\textcolor{blue}{decode and repair}} 

Decode $\textbf{c}'$ and repair it for $\textbf{w}'$ if the constraints are violated.

\tcp{\textcolor{blue}{evaluation}}

Evaluate $\textbf{c}'$ with $\textbf{w}'$.

Set $P' = P' \cup \{c'\}$.
}

\tcp{\textcolor{blue}{selection}}

Select $NP$ solutions to constitute the next population $P$ from $P$ and $P'$.
}
\caption{Algorithm Framework}
\label{ALG1}
\end{algorithm}

In line 1, the first generation population $P$ is randomly initialized in $[0,1]^N$, where $N$ is the number of available assets and $NP$ is the population size. For the fitness evaluation, each solution \textbf{c} in $P$ is decoded by the decoding scheme to a portfolio \textbf{w} (lines 2-4). The main iteration of the algorithm is described in lines 6-14. While the stopping criteria are not met~\cite{liu2018termination}, the new candidate solution is generated with a search operator (line 9). The fitness of each new individual is assessed (line 10), and the new population $P'$ is combined with $P$ for comparison (line 11). Finally, the best individuals are selected by an MOEA selection strategy in each iteration as the next population $P$.

\section{Experiment Study}
\label{section_5}
This section is devoted to the empirical study of the proposed CCS and the new algorithm framework. This section is divided into three parts. Firstly, the instances, the MOEAs, and the corresponding parameters are introduced. Secondly, the quality indicators adopted in this paper are presented. Finally, the MOEAs with CCS (CCS/MOEAs), MODEwAwL~\cite{2014KhinMODEwawl}, and the MOEAs with DCS (DCS/MOEAs) are compared. Two categories of DCS are employed here. First, a canonical one, MODEwAwL, with one binary vector and one real-valued vector. Second, DCS with two real-valued vectors for DCS/MOEAs. The real-valued DCS
is employed for urging the difference between CCS and DCS as small as possible. Because in this manner, DCS can directly use almost the same decoding, search operators, and repairing as CCS. For example, in this DCS, the first vector is mapped to the selection as CCS does and the second vector is mapped to weights as in Eq.~\ref{decoding_ccs}. Thereafter, operators for generating offsprings are the same as in Section~\ref{search_operators}, except the swap of values will take place not only on the selection (first) vector but also on the corresponding weight (second) vector. This means $c_{i+N}$ will swap with $c_{j+N}$ when $c_i$ swaps with $c_j$. Then the same constraints handling method as in Section~\ref{constraint_handling} is also adopted. Besides, a mathematical programming solver is used to solve the instances from OR-Library.

\subsection{Experimental Settings}
Two benchmarks are employed in this paper. The first one, OR-Library, contains 5 classic instances~\cite{2000TJChangHFCCPO}, and the second one, NGINX, includes 15 instances established recently by the historical stock data from the Yahoo Finance website\footnote{OR-Library is from \url{{http://people.brunel.ac.uk/~mastjjb/jeb/orlib/portinfo.html}}, NGINX is from \url{{http://satt.diegm.uniud.it/projects/portfolio-selection/}} (available at September 1st, 2018). We pack them together, named as Portfolio-Instances on \url{https://github.com/CYLOL2019}. Further, the demos for seven algorithms, named as SEC-CCS, can also be found there.}. Table~\ref{Instances} presents the details of each data set, D1-D5 from OR-Library and D6-D20 from NGINX. In addition, two constraint sets~\cite{2014KhinMODEwawl} are considered as follows:
\begin{itemize}
\item[(i)] Cardinality ${K=10}$, floor $\epsilon_i=0.01$, ceiling $\delta_i=1.0$, pre-assignment ${z_{30}=1}$ and round lot $\tau=0.008.$
\item[(ii)]  Cardinality ${K=15}$, floor $\epsilon_i=0.01$, ceiling $\delta_i=1.0$, pre-assignment ${z_5=1}$ and round lot $\tau=0.008.$
\end{itemize}

Note that if not specified, the simulated experiments in this work are constructed with the \emph{first} constraint set.

The majority of the MOEAs is based on three main frameworks, namely, the decomposition-based framework, the Pareto domination based framework, and the indicator-based framework~\cite{zhou2011multiobjective}. Therefore, three widely-used MOEAs, namely, MOEA/D~\cite{2007QingfuMOEAD}, NSGA-\uppercase\expandafter{\romannumeral2}~\cite{2002DebNSGA2}, and SMS-EMOA~\cite{2006NicolaSMS}, corresponding to the three frameworks, are utilized in this study.

\paragraph{Decomposition based Multi-objective Evolutionary Algorithm (MOEA/D)~\cite{2007QingfuMOEAD}} MOEA/D is based on the decomposition framework. 
It converts the multiobjective problems into scalar optimization problems with aggregation functions. Meanwhile, the whole population is optimized simultaneously when each subproblem uses its neighborhood information.

\paragraph{Non-dominated Sorting Genetic Algorithm (NSGA-\uppercase\expandafter{\romannumeral2})~\cite{2002DebNSGA2}} NSGA-\uppercase\expandafter{\romannumeral2} is a popular Pareto domination based MOEA. It uses the Pareto domination relationship and the crowding distance to differentiate solutions. Thus it can select promising solutions for the next generation. 

\paragraph{Multiobjective selection based on dominated hypervolume (SMS-EMOA)~\cite{2006NicolaSMS}} SMS-EMOA is designed as a steady-state MOEA. It uses the Pareto domination relationship and the hypervolume indicator to select promising solutions for the next generation.

In the subsection of the comparison study, DCS and CCS are incorporated into three MOEAs respectively, and a Learning-Guided Multi-Objective Evolutionary Algorithm (MODEwAwL)~\cite{2014KhinMODEwawl}, a canonical DCS algorithm, is also implemented. The algorithms' parameters are shown in Table~\ref{parametersetting}.
\vspace{-5mm}
\begin{table}\tiny
\centering 
\caption{Twenty benchmark instances considered in this study}
\label{Instances}
\begin{tabular}{llll|llll}\hline
Instance&\multicolumn{1}{c}{Origin}&\multicolumn{1}{c}{Name}&\multicolumn{1}{c}{\#Assets} \vline& Instance&\multicolumn{1}{c}{Origin}&\multicolumn{1}{c}{Name}&\multicolumn{1}{c}{\#Assets}\\\hline\hline
$D1$        &Hong Kong   &Hang Seng           &$31$ & $D11$        &UK      &FTSE ACT250               &$128$\\
$D2$        &Germany      &DAX100             &$85$ & $D12$      &USA      &NASDAQ Bank               &$380$\\
$D3$        &UK         &FTSE 100            &$89$ & $D13$        &USA     &NASDAQ Biotech             &$130$\\
$D4$      &US          &S$\&$P 500            &$98$ & $D14$        &USA    &NASDQ  Computer             &$417$\\
$D5$         &Japan        &Nikkei            &$225$ & $D15$        &USA    &NASDQ  Financial00           &$91$\\
$D6$          &Korea    &KOSPI Composite            &$562$ & $D16$        &USA    &NASDQ  Industrial           &$808$\\
$D7$         &USA      &AMEX Composite           &$1893$ & $D17$         &USA     &NASDQ  Telecom             &$139$\\
$D8$        &USA        &NASDAQ                &$2235$ & $D18$          &USA      &NYSE   US100               &$94$\\
$D9$        &Australia   &All ordinaries            &$264$ & $D19$          &USA      &NYSE   World              &$170$\\
$D10$        &Italy       &MIBTEL                 &$167$ & $D20$         &USA       &S$\&$P    500               &$469$\\
\hline\hline
\end{tabular}
\end{table}

\subsection{Quality Indicators}

Two performance metrics, which are well-known and frequently applied, are introduced in this paper. They are Inverted Generational Distance (IGD)~\cite{2005SierraIGD}, and Inverted Hypervolume (IH)~\cite{zitzler1999multiobjective}. Overall, the IGD and IH are general metrics for multi-objective problems, and they cover consideration of both proximity and diversity.\\

\paragraph{Inverted Generational Distance (IGD)~\cite{2005SierraIGD}} IGD, also known as D-metric~\cite{mavrotas2015improved, 2007QingfuMOEAD}, evaluates the distances between every solution and the true PF. It is given as follows.
\begin{flalign*}
&IGD=\frac{\sum_{i=1}^{|Q|}d_{i}}{|Q|},&
\end{flalign*}
where $Q$ is a set of the representatives of the true PF, and $d_i$ is the shortest Euclidean distance between the $i$-th point of the representatives and the solutions obtained by the optimization algorithms. Note that the true PFs for some instances in NGINX are undiscovered. This is because the number of available assets for them is too large. So the best known unconstrained efficient frontiers (UCEFs), provided together with the considered instances, are adopted instead of the true PFs. Hence all researchers can make consistent comparisons with the consensus references.

\begin{table} 
\tiny
\centering
\caption{The parameter setting for all algorithms.}
\label{parametersetting}
\begin{tabular}{ll|ll}\hline
Common parameters&\multicolumn{1}{c}{}&\multicolumn{1}{l}{Parameters only for MOEA/D}&\multicolumn{1}{c}{}\\\hline
Population size $(NP)$    &100    &Neighborhood size     &10\\
Number of generations    &1000    &All solutions as the neighbors probability $p_{\delta}$ &0.1 \\
Scaling factor $F$    &0.5    &Replacement size $T_r$ & 2 \\
Crossover probability $CR$    &0.9 \\
Index parameter $\eta_m$     &20    \\
Polynomial mutation probability $p_m$ &$\frac{1}{NP}$\\

\hline\hline
\end{tabular}
\end{table}

\paragraph{Inverted Hypervolume(IH)} IH is the inverted version of Hypervolume (HV)~\cite{zitzler1999multiobjective}. HV, also known as the size of dominated space, is a quality indicator that rewards the convergence toward the PF as well as the distribution of the representative points along the front. It normalizes the objective space and measures the volume of space, which is bounded by the obtained \emph{efficient} solutions and a preference point $r$. For each obtained solution $i\in Q$, a hypercube $hc_i$ from solution $i$, and the reference point $r$ is measured. Generally, higher values of HV are preferable. However, to apply a consistent comparison as IGD, IH is defined by
\begin{flalign*}
&IH = {\rm volume}(\cup_{i=1}^{|Q|}(hc_r-hc_i)),&
\end{flalign*}
where $hc_r$ is the hypercube among the reference point and representatives. Hence, lower values are better concerning this definition. In addition, the reference point is set to $[1.2, 1.2]$ while all the solutions are normalized for minimizing $f_1$ and $-f_2$.
\begin{figure}[htbp]

\graphicspath{{figs/}}
\centering
\tiny
\begin{minipage}{0.45\linewidth}
\label{fig_AUGMOEA:a}
\centerline{\includegraphics[width=1\columnwidth]{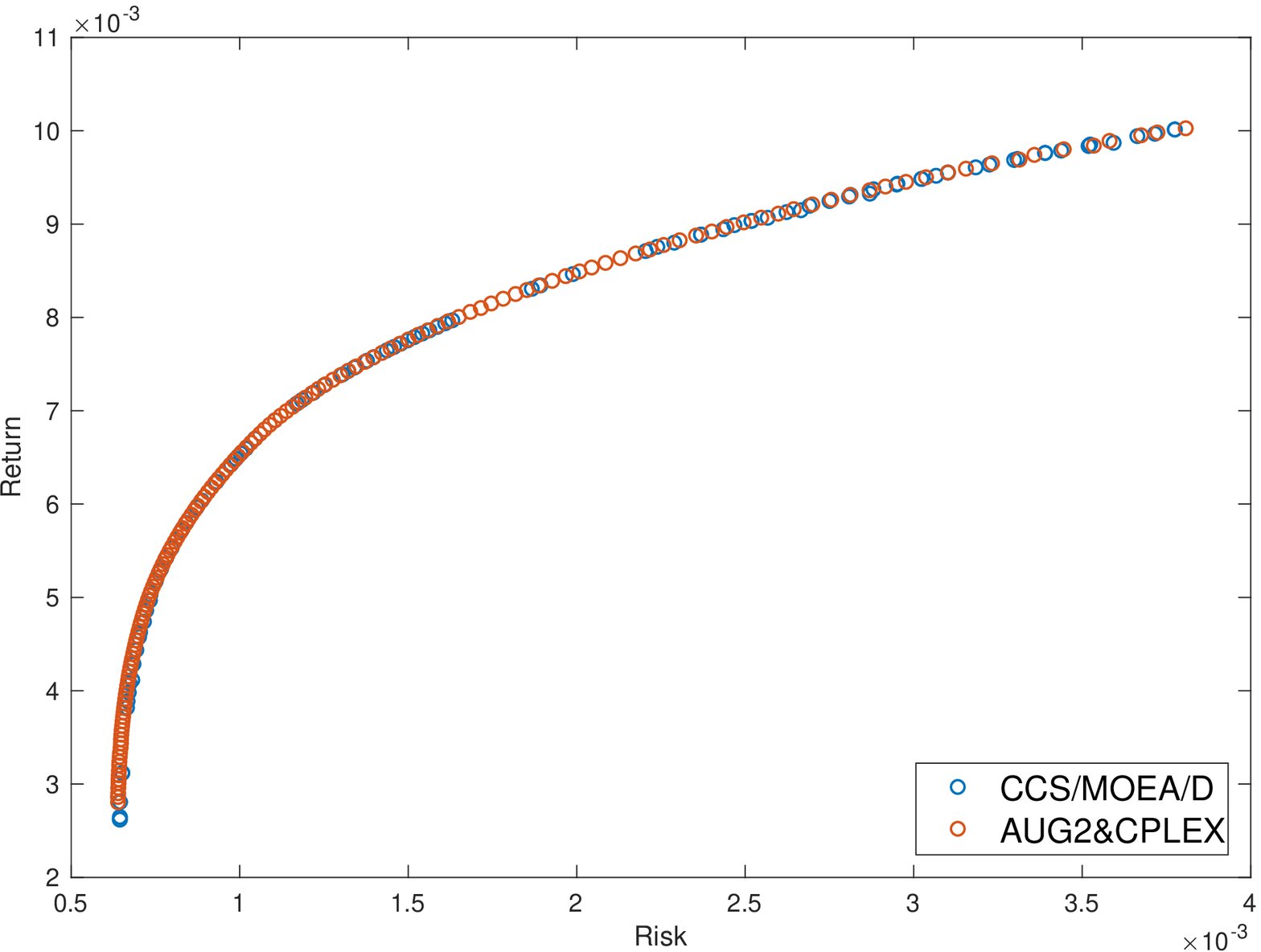}}
\centerline{(a) D1}
\end{minipage}
\begin{minipage}{0.45\linewidth}
\label{fig_AUGMOEA:b}
\centerline{\includegraphics[width=1\columnwidth]{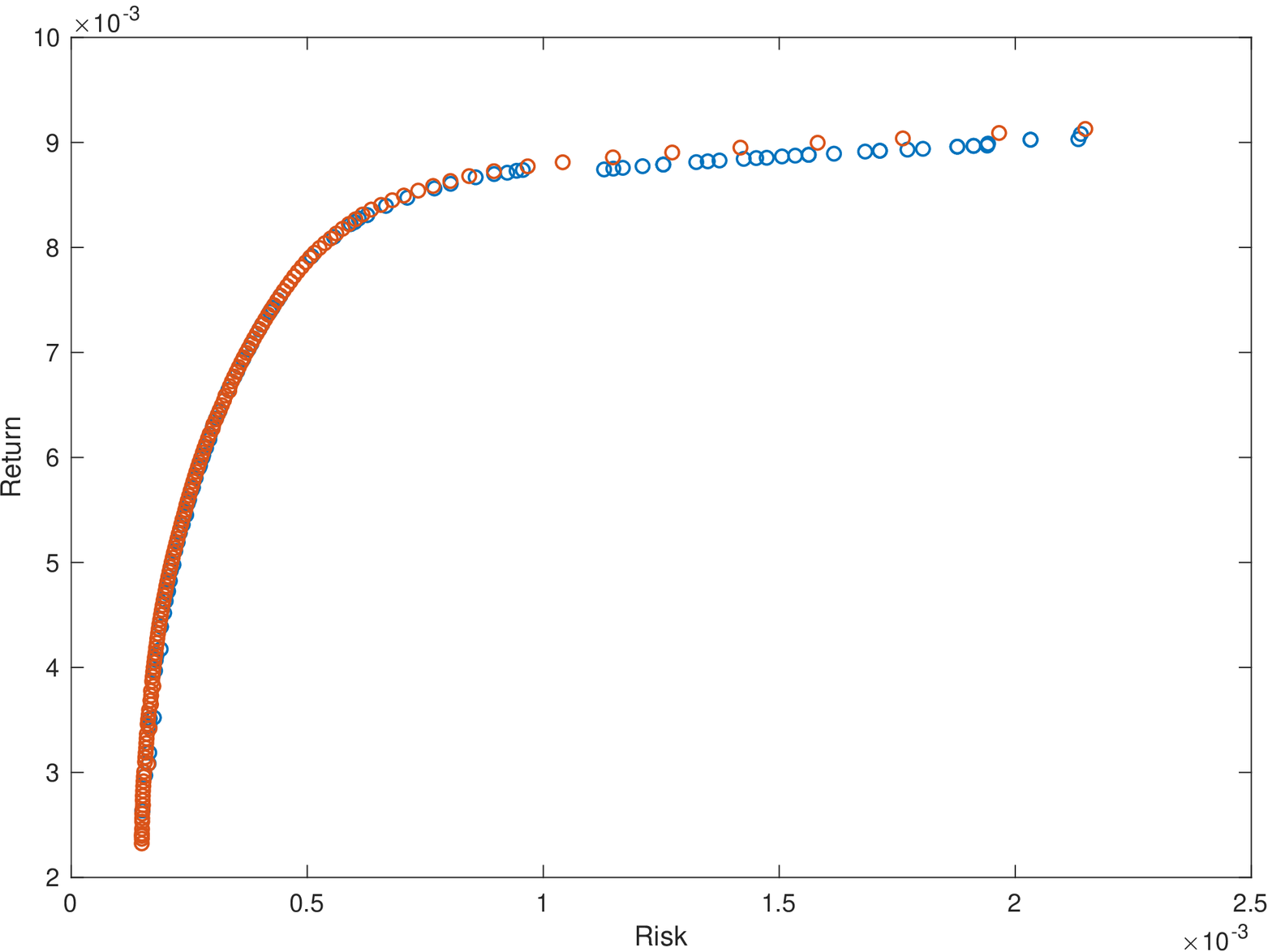}}
\centerline{(b) D2}
\end{minipage}
\begin{minipage}{0.45\linewidth}
\label{fig_AUGMOEA:c}
\centerline{\includegraphics[width=1\columnwidth]{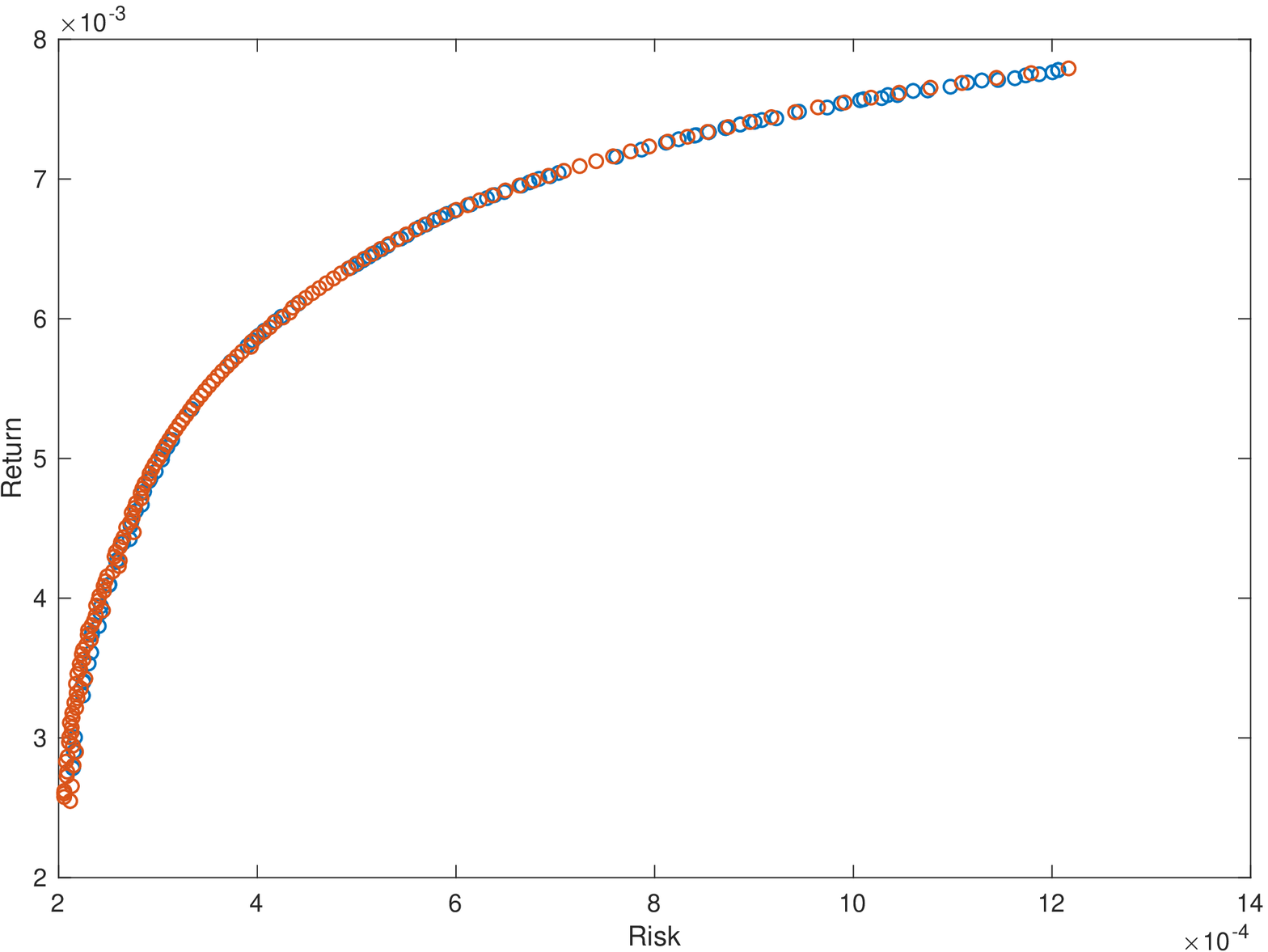}}
\centerline{(c) D3}
\end{minipage}
\begin{minipage}{0.45\linewidth}
\label{fig_AUGMOEA:d}
\centerline{\includegraphics[width=1\columnwidth]{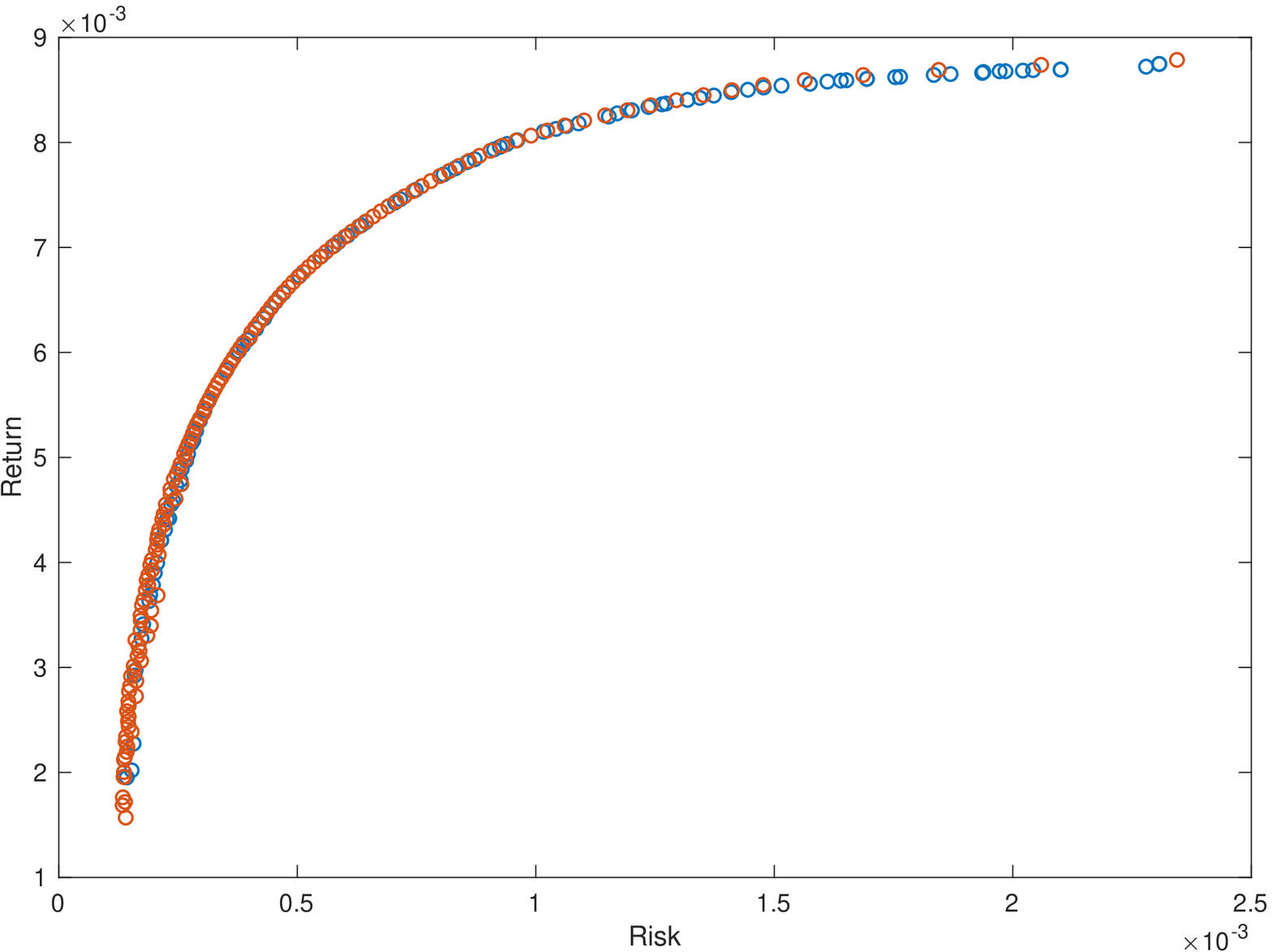}}
\centerline{(d) D4}
\end{minipage}
\begin{minipage}{0.45\linewidth}
\label{fig_AUGMOEA:e}
\centerline{\includegraphics[width=1\columnwidth]{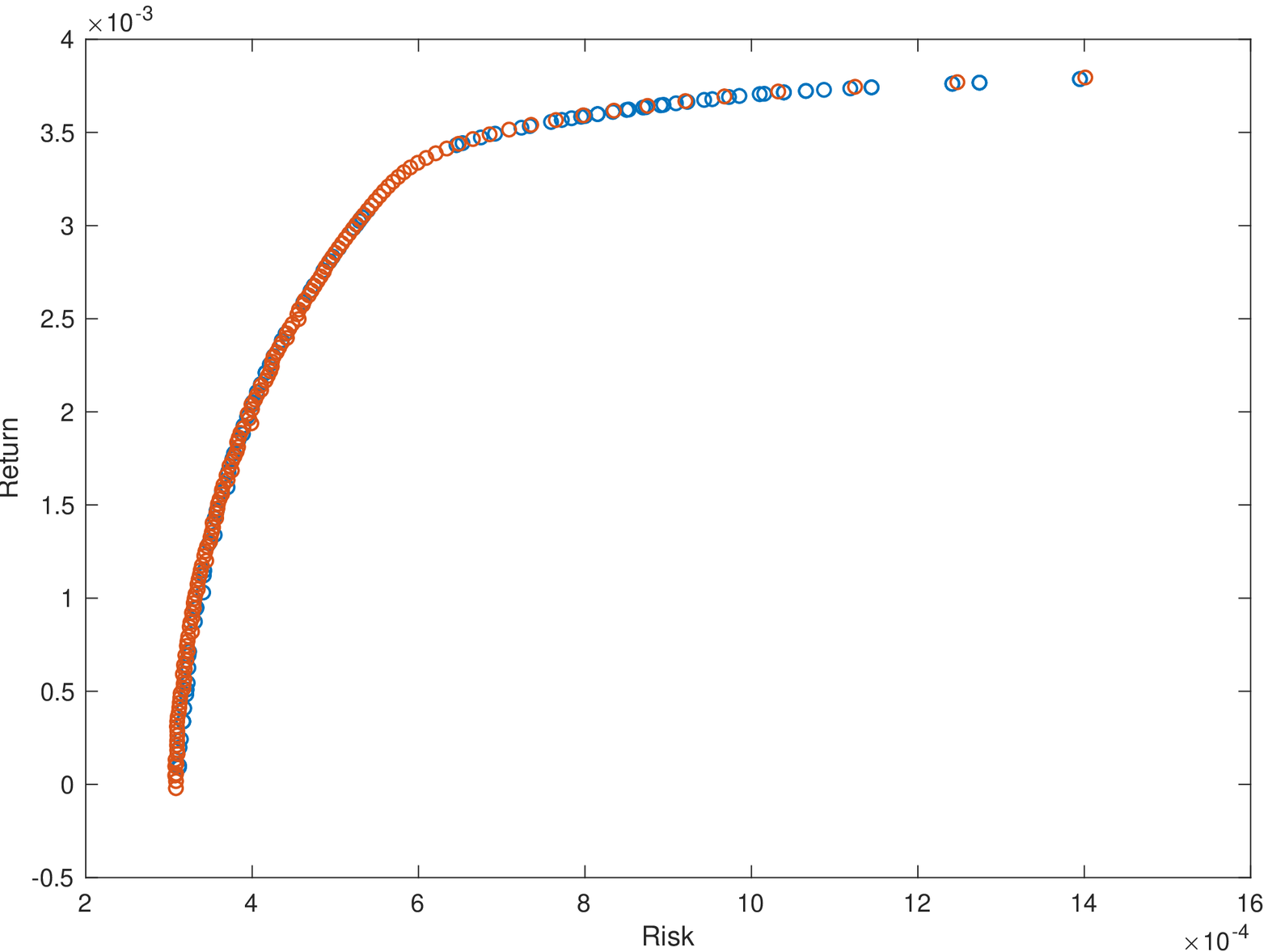}}
\centerline{(e) D5}
\end{minipage}
 \caption{The efficient fronts obtained by AUGMECON2\&CPLEX and CCS/MOEA/D on OR-Library. The efficient fronts of CCS/MOEA/D are the best ones concerning IGD.} 
 \label{fig_AUGMOEA}
\end{figure}
\begin{table*}
\tiny
 \caption{Performance of 7 algorithms on OR-Library concerning IGD. } 
\label{OR_IGD_Constraints1}
\begin{tabular}{*{11}{c}}
\hline
\multicolumn{2}{c}{Algo.}&\multicolumn{1}{c}{A}&\multicolumn{1}{c}{B}&\multicolumn{1}{c}{C}&\multicolumn{1}{c}{D}&\multicolumn{1}{c}{E}&\multicolumn{1}{c}{F}&\multicolumn{1}{c}{G}\\\hline\hline
\multirow{2}{*}{D1}
&Mean	&1.61e-02[6]	&6.90e-03[2]	&9.68e-02[7]	&7.20e-03[3]	&1.43e-02[5]	&\cellcolor{gray25}5.79e-03[1]	&1.23e-02[4]	\\
&Std	&4.99e-03	&5.45e-04	&4.25e-02	&8.58e-04	&1.96e-02	&7.25e-04	&2.33e-03	\\\hline
\multirow{2}{*}{D2}
&Mean	&1.11e-02[5]	&1.03e-02[4]	&2.98e-02[7]	&8.76e-03[3]	&6.02e-03[2]	&\cellcolor{gray25}4.70e-03[1]	&2.08e-02[6]	\\
&Std	&2.44e-03	&2.28e-03	&9.45e-03	&7.24e-04	&3.20e-03	&2.72e-04	&7.71e-03	\\\hline
\multirow{2}{*}{D3}
&Mean	&1.19e-02[5]	&7.66e-03[2]	&4.26e-02[7]	&9.63e-03[4]	&8.20e-03[3]	&\cellcolor{gray25}5.31e-03[1]	&2.43e-02[6]	\\
&Std	&7.16e-03	&1.15e-03	&1.82e-02	&2.71e-03	&4.01e-03	&1.27e-03	&7.30e-03	\\\hline
\multirow{2}{*}{D4}
&Mean	&1.91e-02[5]	&7.01e-03[2]	&3.83e-02[7]	&1.13e-02[4]	&1.07e-02[3]	&\cellcolor{gray25}6.22e-03[1]	&2.86e-02[6]	\\
&Std	&9.71e-03	&6.97e-04	&9.06e-03	&2.06e-03	&7.80e-03	&3.58e-03	&7.07e-03	\\\hline
\multirow{2}{*}{D5}
&Mean	&8.95e-02[5]	&1.25e-02[3]	&9.16e-02[6]	&\cellcolor{gray25}9.71e-03[1]	&1.28e-01[7]	&1.16e-02[2]	&4.04e-02[4]	\\
&Std	&2.60e-02	&2.35e-03	&1.71e-02	&3.26e-03	&5.20e-02	&5.96e-03	&7.70e-03	\\\hline
\multicolumn{2}{c}{MeanRank}&5.2 	&2.6 	&6.8 	&3.0 	&4.0 	&1.2 	&5.2 	\\\hline
\hline
\multicolumn{3}{c}{+/-/=$^{1}$}&\multicolumn{2}{|c}{0/4/1	(A\&B)}&\multicolumn{2}{|c}{0/5/0	(C\&D)}&\multicolumn{2}{|c}{0/5/0	(E\&F)}	\\\hline
\multicolumn{2}{c}{+/-/=$^{2}$}&\multicolumn{1}{|c}{2/3/0}&\multicolumn{1}{|c}{0/5/0}&\multicolumn{1}{|c}{5/0/0}&\multicolumn{1}{|c}{0/5/0}&\multicolumn{1}{|c}{2/3/0}&\multicolumn{1}{|c}{0/5/0}&\multicolumn{1}{|c}{-}\\
\hline
\end{tabular}
\end{table*}
\begin{table*}
\tiny
 \caption{Performance of 7 algorithms on OR-Library concerning IH. } 
\label{OR_IH_Constraints1}
\begin{tabular}{*{11}{c}}
\hline
\multicolumn{2}{c}{Algo.}&\multicolumn{1}{c}{A}&\multicolumn{1}{c}{B}&\multicolumn{1}{c}{C}&\multicolumn{1}{c}{D}&\multicolumn{1}{c}{E}&\multicolumn{1}{c}{F}&\multicolumn{1}{c}{G}\\\hline\hline
\multirow{2}{*}{D1}
&Mean	&3.09e-02[6]	&7.85e-03[2]	&2.54e-01[7]	&9.33e-03[3]	&2.34e-02[5]	&\cellcolor{gray25}4.94e-03[1]	&1.48e-02[4]	\\
&Std	&1.07e-02	&8.01e-04	&9.51e-02	&1.21e-03	&4.55e-02	&8.75e-04	&2.34e-03	\\\hline
\multirow{2}{*}{D2}
&Mean	&1.65e-02[5]	&1.33e-02[4]	&6.25e-02[7]	&1.22e-02[3]	&7.14e-03[2]	&\cellcolor{gray25}3.78e-03[1]	&3.02e-02[6]	\\
&Std	&4.51e-03	&3.82e-03	&3.38e-02	&1.79e-03	&4.15e-03	&8.88e-04	&1.99e-02	\\\hline
\multirow{2}{*}{D3}
&Mean	&2.09e-02[5]	&1.12e-02[2]	&8.24e-02[7]	&1.57e-02[4]	&1.35e-02[3]	&\cellcolor{gray25}7.47e-03[1]	&3.65e-02[6]	\\
&Std	&1.34e-02	&2.86e-03	&3.57e-02	&5.17e-03	&6.99e-03	&2.52e-03	&1.39e-02	\\\hline
\multirow{2}{*}{D4}
&Mean	&3.38e-02[5]	&1.15e-02[2]	&7.67e-02[7]	&1.68e-02[4]	&1.65e-02[3]	&\cellcolor{gray25}8.53e-03[1]	&4.59e-02[6]	\\
&Std	&2.09e-02	&2.12e-03	&3.07e-02	&3.26e-03	&9.89e-03	&6.14e-03	&1.60e-02	\\\hline
\multirow{2}{*}{D5}
&Mean	&2.03e-01[5]	&1.77e-02[2]	&2.05e-01[6]	&\cellcolor{gray25}1.76e-02[1]	&2.77e-01[7]	&2.44e-02[3]	&5.30e-02[4]	\\
&Std	&5.05e-02	&3.46e-03	&5.22e-02	&7.74e-03	&1.19e-01	&1.74e-02	&9.86e-03	\\\hline
\multicolumn{2}{c}{MeanRank}&5.2 	&2.4 	&6.8 	&3.0 	&4.0 	&1.4 	&5.2 	\\\hline
\hline
\multicolumn{3}{c}{+/-/=$^{1}$}&\multicolumn{2}{|c}{0/5/0	(A\&B)}&\multicolumn{2}{|c}{0/5/0	(C\&D)}&\multicolumn{2}{|c}{0/5/0	(E\&F)}	\\\hline
\multicolumn{2}{c}{+/-/=$^{2}$}&\multicolumn{1}{|c}{2/3/0}&\multicolumn{1}{|c}{0/5/0}&\multicolumn{1}{|c}{5/0/0}&\multicolumn{1}{|c}{0/5/0}&\multicolumn{1}{|c}{2/3/0}&\multicolumn{1}{|c}{0/5/0}&\multicolumn{1}{|c}{-}\\
\hline
\end{tabular}
\end{table*}

\subsection{Exact Method}
In the field of optimization, heuristic and exact methods are two equally significant components. Hence it is worth making a comparison between these two kinds of methods. Admittedly, there are probably many exact methods for these problems, nonetheless, we can not cover all of them due to space constraints. AUGMECON2~\cite{mavrotas2013improved}, a state-of-the-art multi-objective programming approach, integrated with CPLEX is implemented here. AUGMECON~\cite{mavrotas2009effective}, the basis for AUGMECON2, proposes an effective method to make a better payoff table, which accelerates the search by avoiding the production of weakly Pareto optimal solutions. Then a faster search approach is provided by AUGMECON2, it utilizes a jumping strategy on the gridded objective values to accelerate the iterative search. This integrated method is only compared with CCS/MOEA/D, since the running times and performance of them are similar at the macro level. Mere data sets from OR-library, D1-D5, are used here because the results obtained from the exact method are similar on small scale problems, no more than 225 assets. Meanwhile, we can not get solutions after executing the exact method in some instances, containing more than 225 assets, from NGINX for 7 days.

Note that the number of the grid is set to 100, the same as the population size of the MOEAs. The results are shown in Fig.~\ref{fig_AUGMOEA}. For all these 5 instances, the exact method can always get the Pareto optima. This is most obvious in Fig. 6(b), the red points with the highest return and lowest risk are obviously far from the blues ones, and most of the blue points with risk more than 1e-3 are dominated by the red points. Moreover, the running times of the exact method on all instances are from 10 seconds to 1 minute, while they are from 5 to 15 minutes for the MOEAs based on the number of available assets. The exact method is apparently faster than the MOEA. Nonetheless, distributions of the blue points are better than the red ones. For example, in Fig. 6(b), blue points distribute uniformly while the red ones scattering with high risk, which is due to the uniform gridding strategy for objective values.
 
These results imply the superiorities of the exact method are (i) faster, and (ii) higher accuracy, while the MOEAs, CCS/MOEA/D here, are more general since they (i) can get uniform solutions without special settings and (ii) can solve large portfolio problems, like instances involving more than 225 assets.

\begin{figure}[htbp]
\centering
\graphicspath{{figs/}}
\subfigure{\includegraphics[width=0.32\columnwidth]{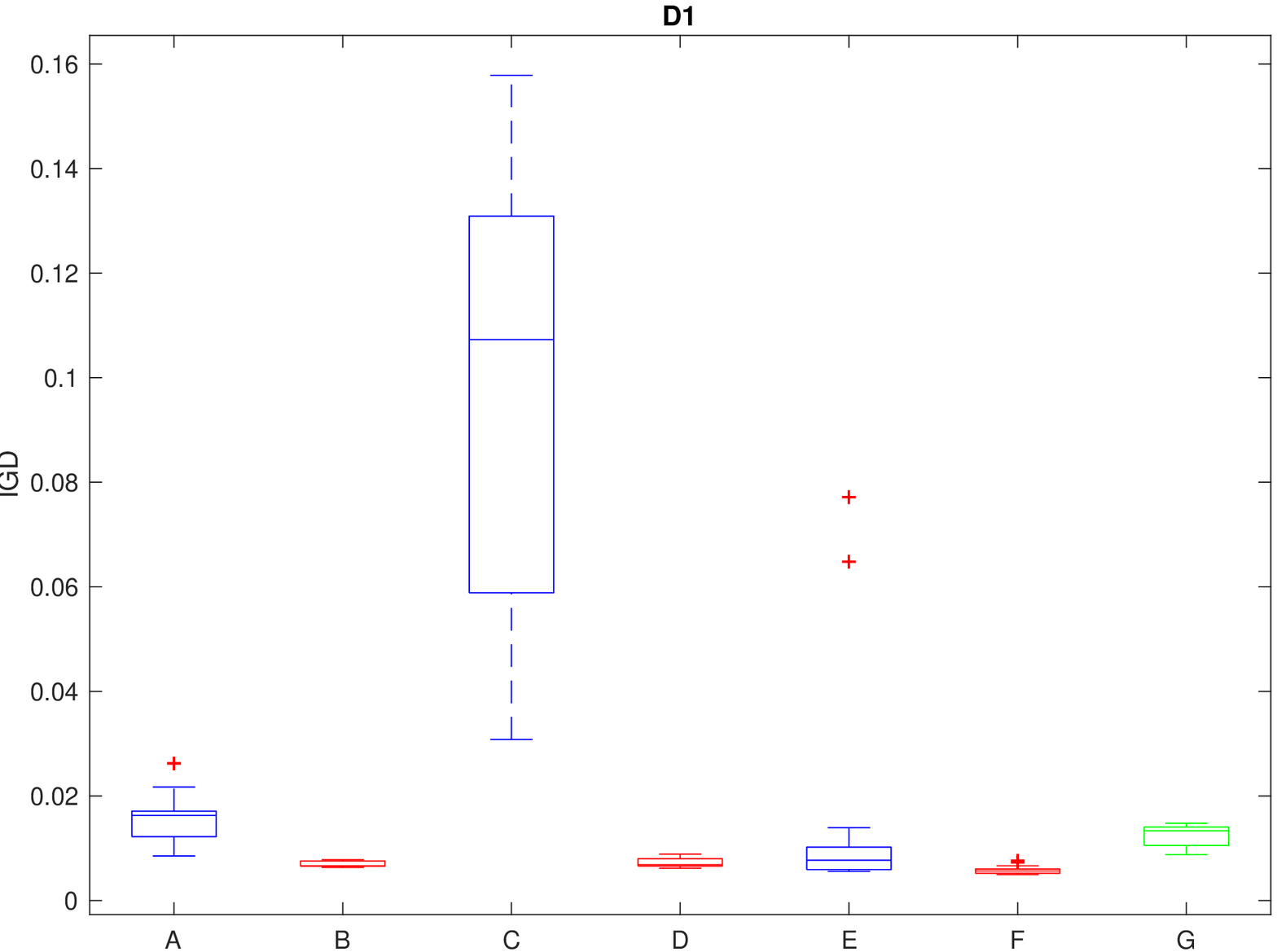}}
\subfigure{\includegraphics[ width=0.32\columnwidth]{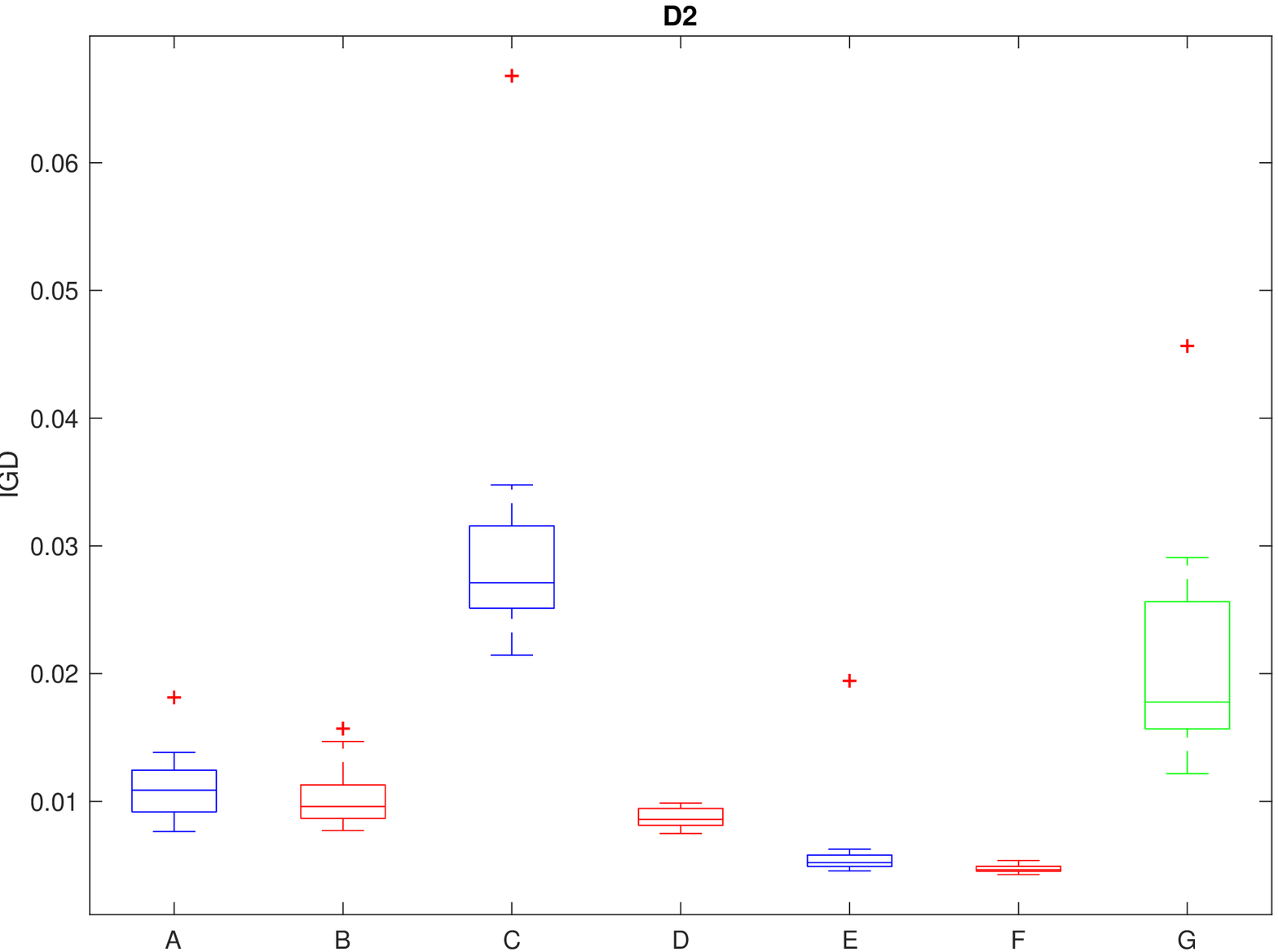}}
\subfigure{\includegraphics[ width=0.32\columnwidth]{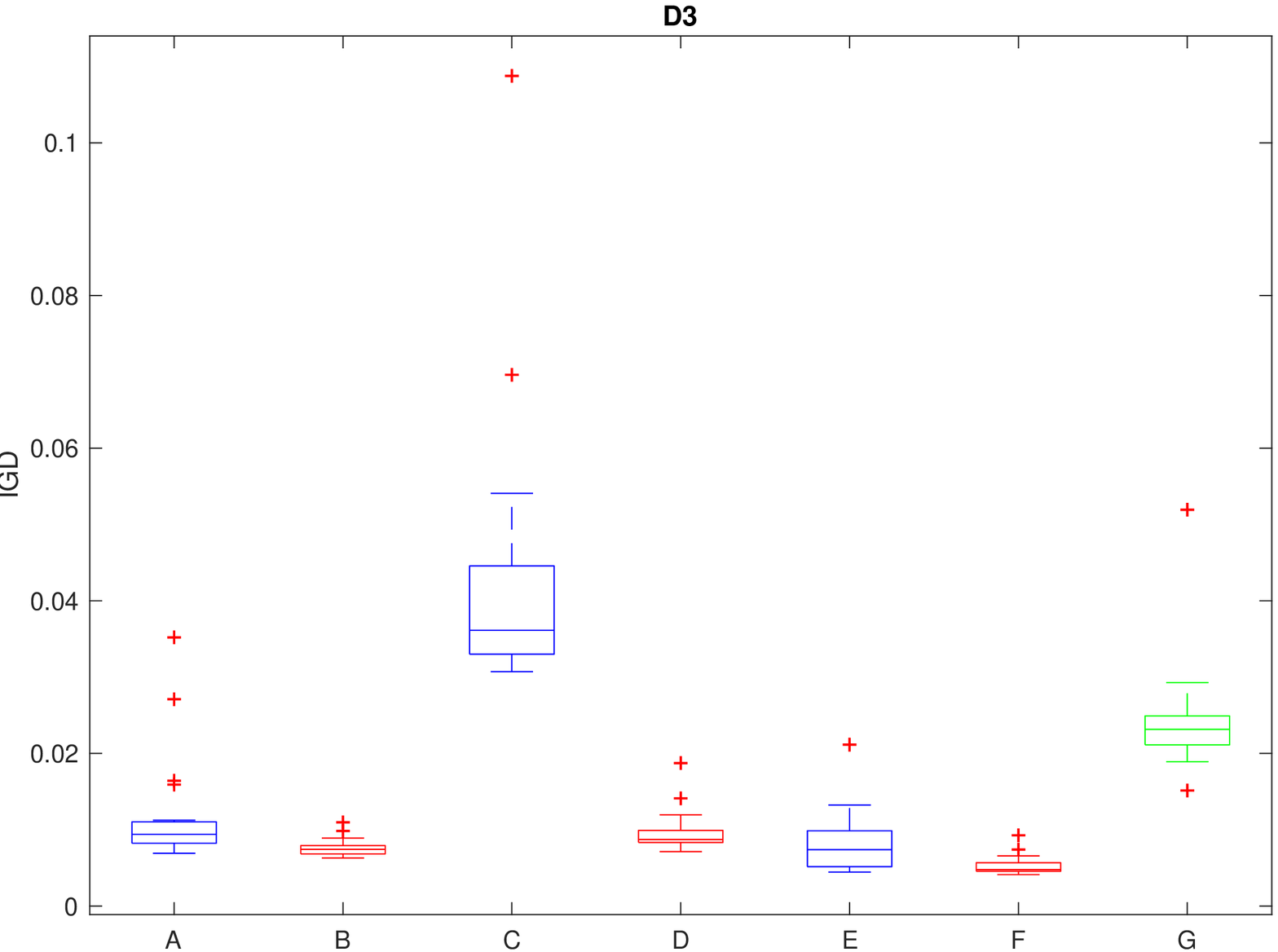}}\\
\subfigure{\includegraphics[ width=0.32\columnwidth]{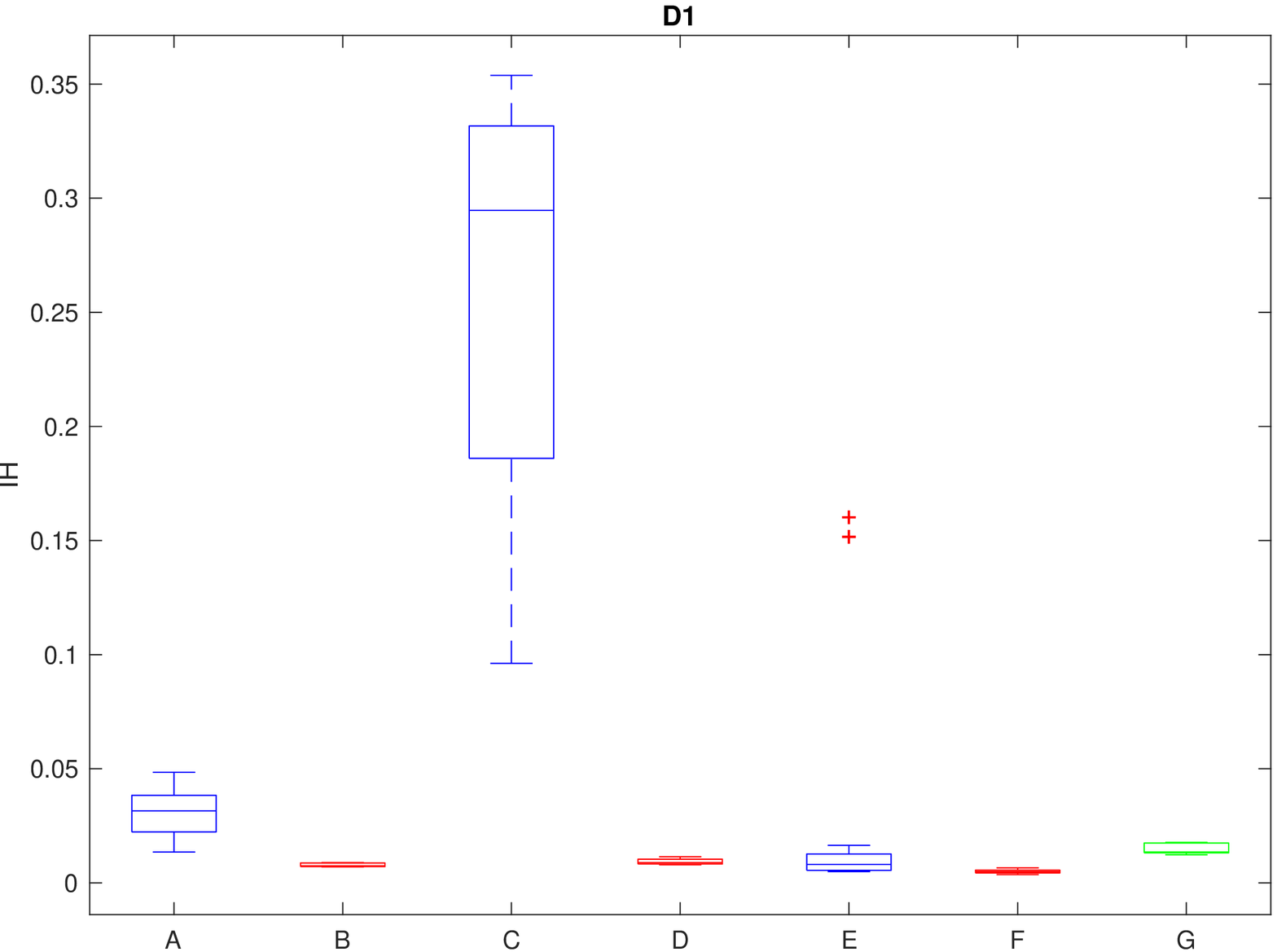}}
\subfigure{\includegraphics[ width=0.32\columnwidth]{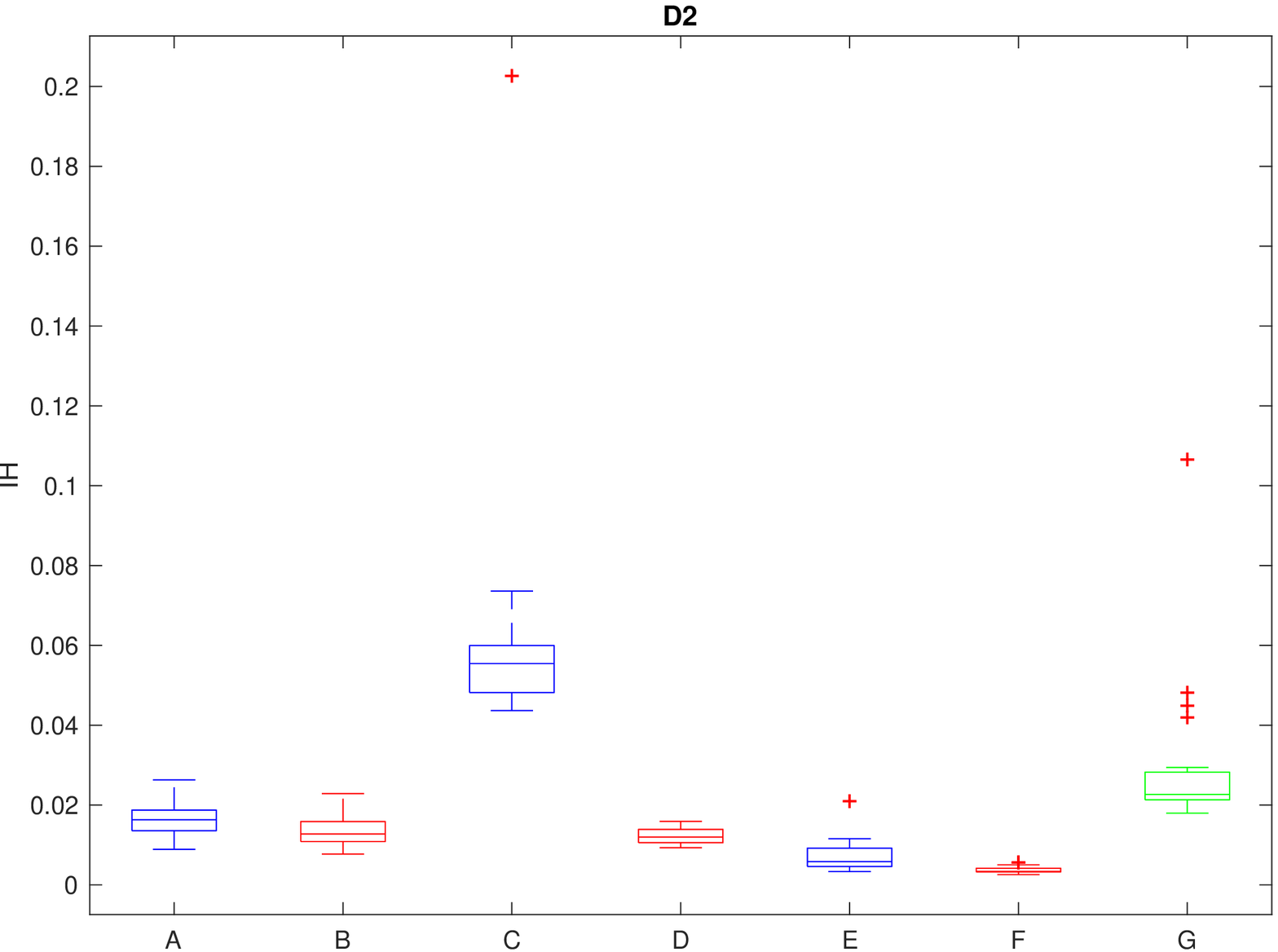}}
\subfigure{\includegraphics[ width=0.32\columnwidth]{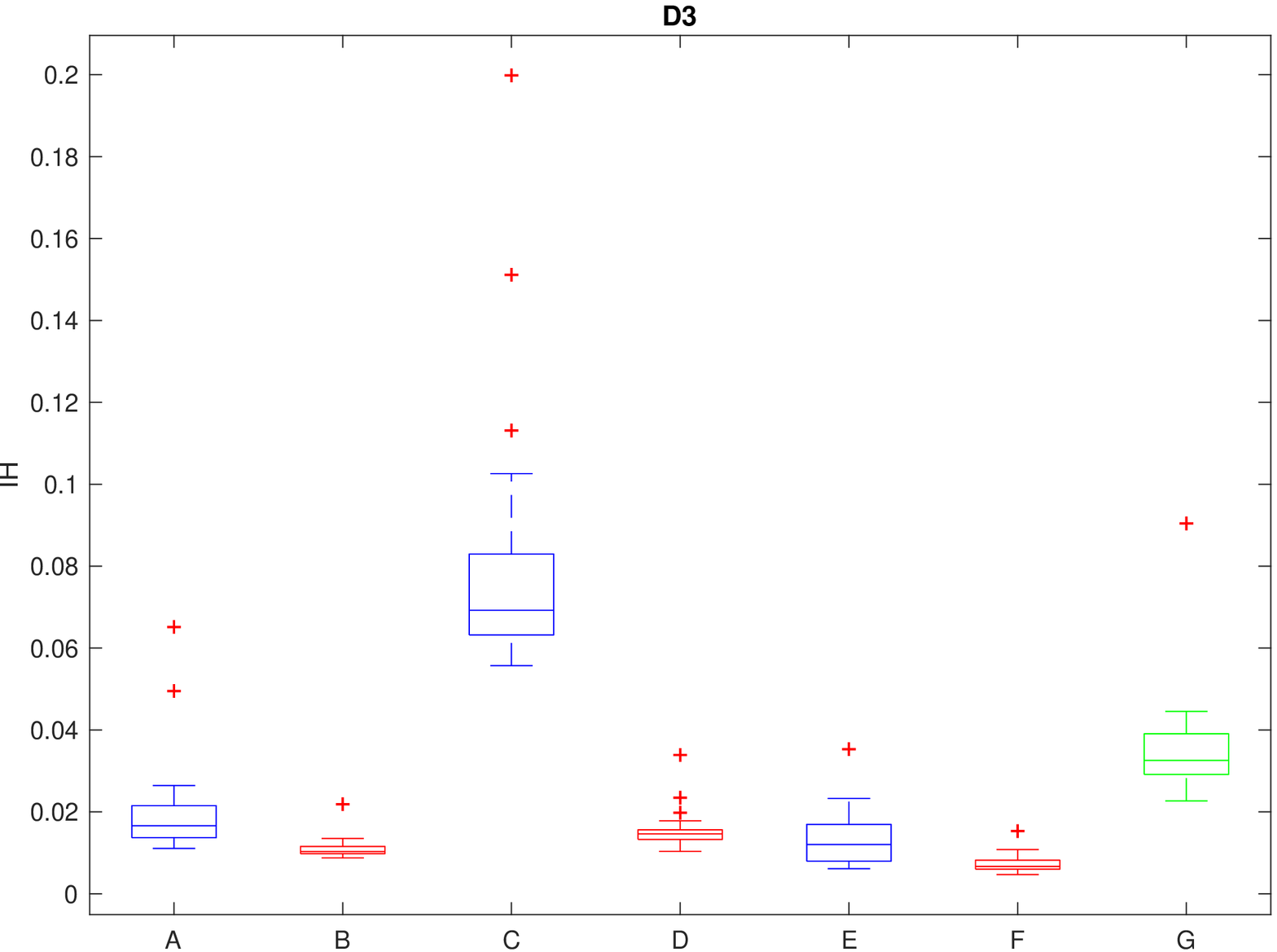}}\\
\subfigure{\includegraphics[ width=0.32\columnwidth]{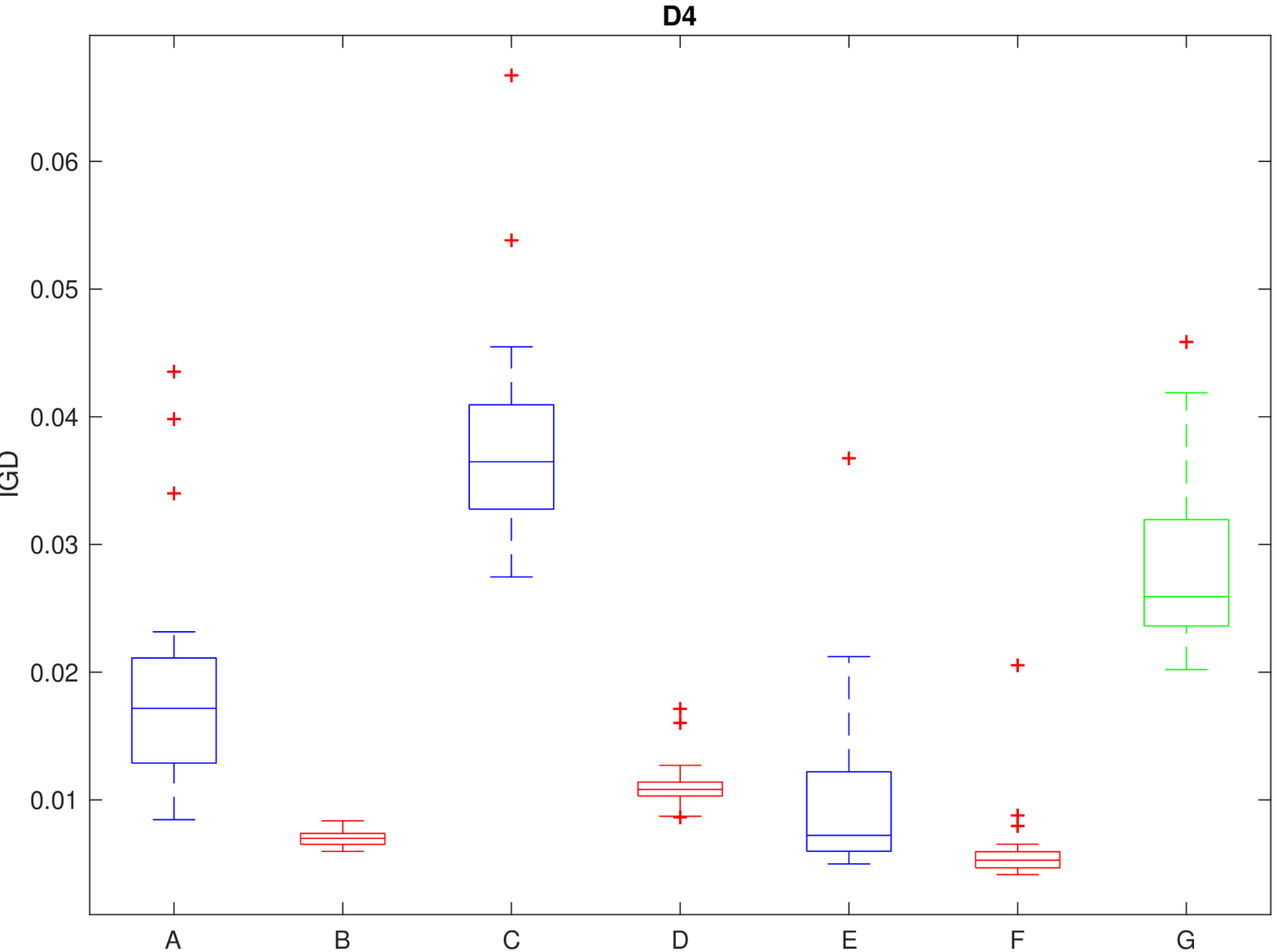}}
\subfigure{\includegraphics[ width=0.32\columnwidth]{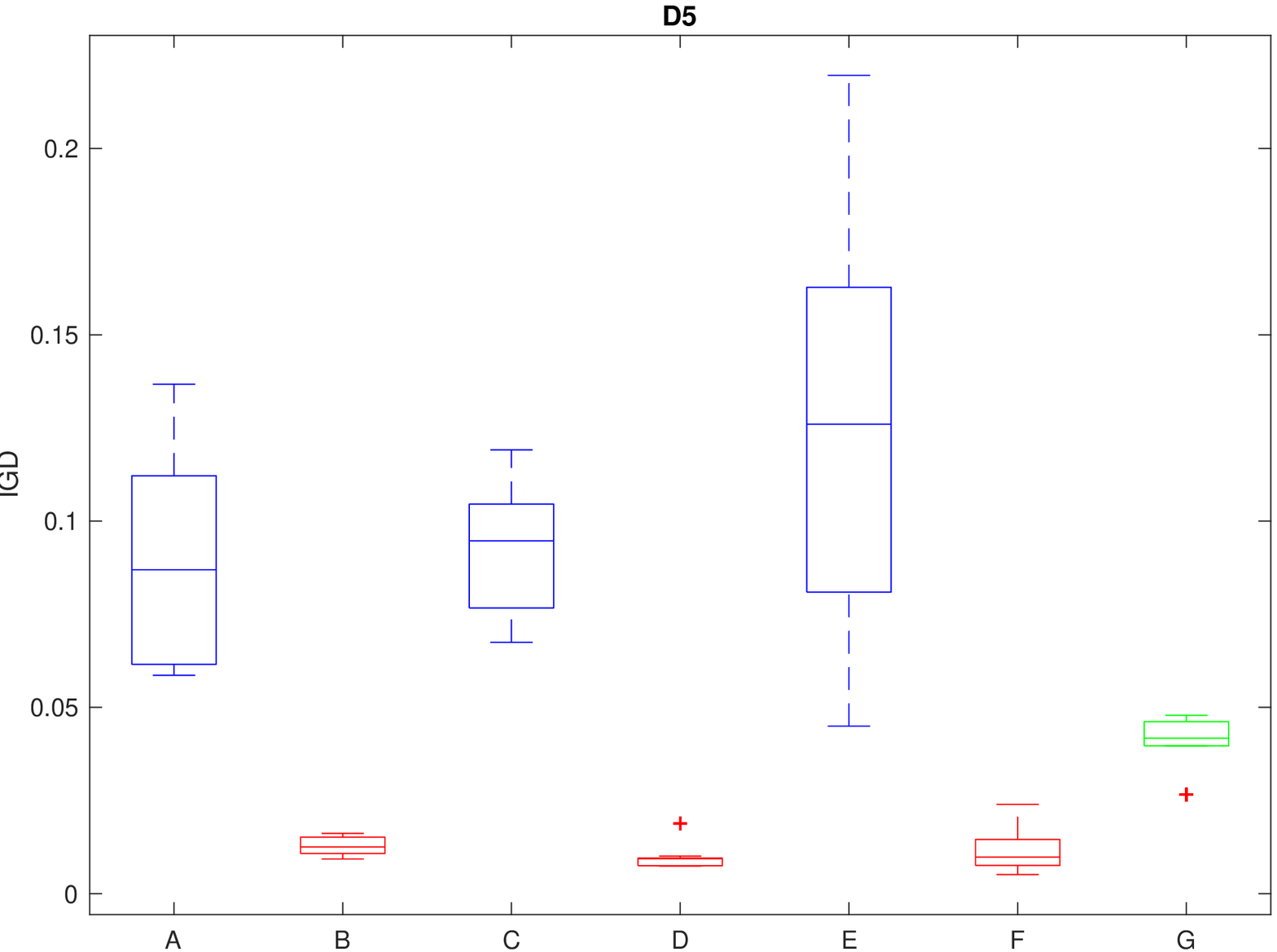}}\\
\subfigure{\includegraphics[ width=0.32\columnwidth]{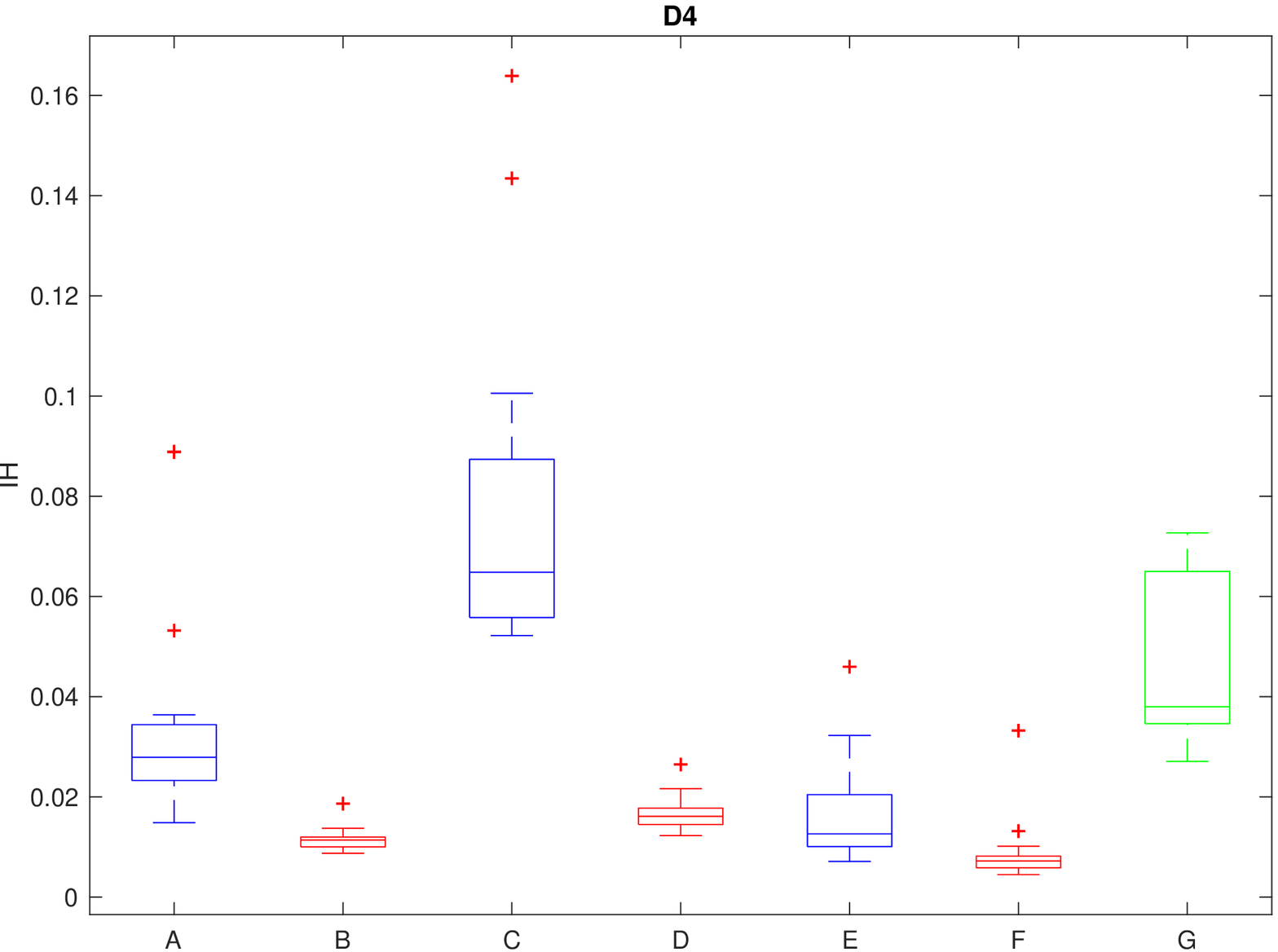}}
\subfigure{\includegraphics[ width=0.32\columnwidth]{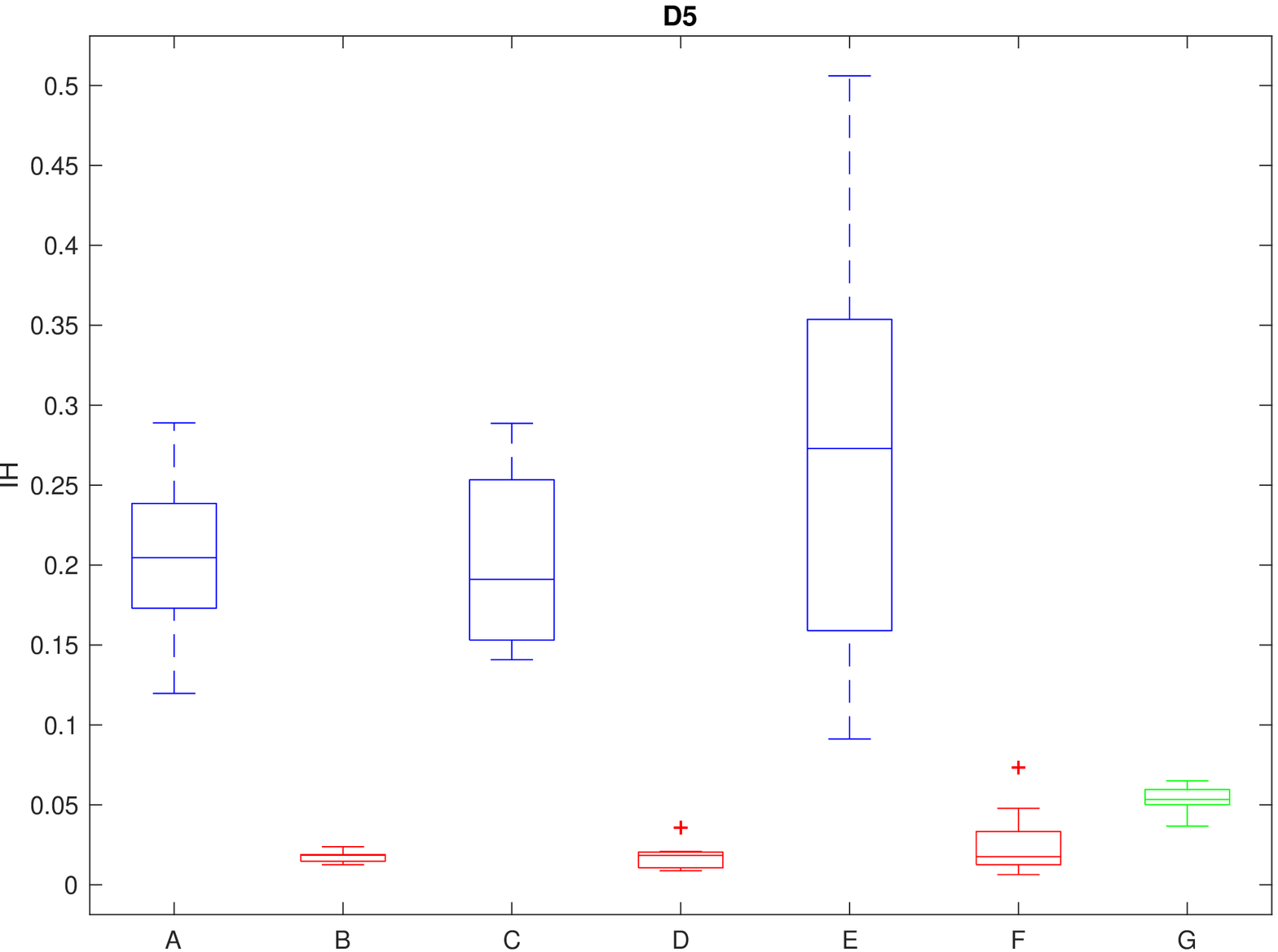}}

 \caption{Performance of 7 algorithms in terms of IGD and IH on OR-Library.}
 \label{port_box}
\end{figure}
\subsection{Comparison Study among MOEAs}

The comparison study among MOEAs aims to answer the following questions.
\begin{itemize}
\item What is the performance of MOEAs with CCS?
\item What is the contribution of CCS?
\end{itemize}
To answer the first question, CCS/MOEAs are compared with a state-of-the-art algorithm, i.e., MODEwAwl. To answer the second question, CCS/MOEAs are compared to DCS/MOEAs~\cite{chen2017evolutionary}.

All seven algorithms are executed for 20 independent runs. As for the performance tables and figures, first, symbols A-G denote the algorithms DCS/MOEA/D, CCS/MOEA/D, DCS/NSGA-\uppercase\expandafter{\romannumeral2}, CCS/NSGA-\uppercase\expandafter{\romannumeral2}, DCS/SMS, CCS/SMS and MODEwAwL respectively. Second, the rank and mean rank of each algorithm on all instances are listed, especially, lower rank means better performance. Third, the best result of each instance is remarked in gray. Finally, the symbol ``+/-/=" indicates better, worse, and equal performance relationship between paired algorithms in terms of Wilcoxon rank-sum test\footnote{Wilcoxon-rank-sum test at a 5\%significant level is adopted in this paper}~\cite{hurlbert1984pseudoreplication} while the symbols with superscript $1$ and $2$ correspond to a one-paired and a six-paired comparison respectively.

Firstly, the comparisons are made on OR-Library, meanwhile, the representatives of the true PFs are obtained by the mathematic method with 1000 grids. As for CCS/MOEAs and MODEwAwL, in terms of IGD, Table~\ref{OR_IGD_Constraints1} shows that all the best mean ranks are reached by the CCS/MOEAs, and MODEwAwL is statistically worse than the three CCS/MOEAs concerning Wilcoxon rank-sum test, all of them are `0/5/0'. Regarding the CCS/MOEAs and the DCS/MOEAs, it shows that all the CCS/MOEAs are better than the corresponding DCS/MOEAs. For example, CCS/MOEA/D is better than DCS/MOEA/D on all 5 instances concerning the mean IGD. Furthermore, the mean rank of CCS/MOEA/D is 2.6 while it is 5.2 for DCS/MOEA/D. As for the Wilcoxon rank-sum test, CCS/MOEA/D is statistically better than DCS/MOEA/D on 4 and similar on 1 dataset. In Table~\ref{OR_IH_Constraints1}, it shows the results in terms of IH, the results are consistent with Table~\ref{OR_IGD_Constraints1}. To be specific, CCS/MOEAs are the top three algorithms out of seven concerning the mean rank, they are also statistically better than MODEwAwL, and none of the CCS/MOEAs is worse than the corresponding DCS/MOEAs.

Furthermore, the results about IGD and IH of all the seven algorithms on OR-Library are presented in Fig.~\ref{port_box}. CCS/MOEAs (red boxes) are better than the corresponding DCS/MOEAs (blue boxes), and they are also better than MODEwAwL in terms of both IGD and IH. Moreover, all the CCS/MOEAs perform similarly while the performance of the DCS/MOEAs fluctuates with different datasets.

To conclude, the results above consistently indicate two significant points. (i) The CCS/MOEAs outperform the DCS/MOEAs and MODEwAwL on the given instances in the quality of the obtained results, and (ii) furthermore, CCS/MOEAs are more robust because the MOEAs perform similarly when using CCS.
\begin{table*}
\tiny
\centering
 \caption{Performance of 7 algorithms on NGINX concerning IGD. } 
\label{NGINX_IGD_Constraints1}
\begin{tabular}{*{11}{c}}
\hline
\multicolumn{2}{c}{Algo.}&\multicolumn{1}{c}{A}&\multicolumn{1}{c}{B}&\multicolumn{1}{c}{C}&\multicolumn{1}{c}{D}&\multicolumn{1}{c}{E}&\multicolumn{1}{c}{F}&\multicolumn{1}{c}{G}\\\hline
\multicolumn{2}{c}{MeanRank}&4.1 	&2.6 	&6.7 	&3.3 	&4.5 	&3.1 	&3.7 	\\\hline

\multicolumn{3}{c}{+/-/=$^{1}$}&\multicolumn{2}{|c}{0/6/9	(A\&B)}&\multicolumn{2}{|c}{0/13/2	(C\&D)}&\multicolumn{2}{|c}{0/6/9	(E\&F)}	\\\hline
\multicolumn{2}{c}{+/-/=$^{2}$}&\multicolumn{1}{|c}{6/3/6}&\multicolumn{1}{|c}{5/5/5}&\multicolumn{1}{|c}{14/0/1}&\multicolumn{1}{|c}{6/6/3}&\multicolumn{1}{|c}{5/4/6}&\multicolumn{1}{|c}{2/6/7}&\multicolumn{1}{|c}{-}\\
\hline
\end{tabular}
\end{table*}
\begin{table*}
\tiny
\centering
 \caption{Performance of 7 algorithms on NGINX concerning IH. } 
\label{NGINX_IH_Constraints1}
\begin{tabular}{*{11}{c}}
\hline
\multicolumn{2}{c}{Algo.}&\multicolumn{1}{c}{A}&\multicolumn{1}{c}{B}&\multicolumn{1}{c}{C}&\multicolumn{1}{c}{D}&\multicolumn{1}{c}{E}&\multicolumn{1}{c}{F}&\multicolumn{1}{c}{G}\\\hline
\multicolumn{2}{c}{MeanRank}&4.7 	&3.0 	&6.9 	&3.4 	&3.9 	&1.5 	&4.7 	\\
\hline
\multicolumn{3}{c}{+/-/=$^{1}$}&\multicolumn{2}{|c}{0/7/8	(A\&B)}&\multicolumn{2}{|c}{0/15/0	(C\&D)}&\multicolumn{2}{|c}{0/14/1	(E\&F)}	\\\hline
\multicolumn{2}{c}{+/-/=$^{2}$}&\multicolumn{1}{|c}{5/6/4}&\multicolumn{1}{|c}{3/11/1}&\multicolumn{1}{|c}{15/0/0}&\multicolumn{1}{|c}{3/10/2}&\multicolumn{1}{|c}{5/9/1}&\multicolumn{1}{|c}{1/11/3}&\multicolumn{1}{|c}{-}\\
\hline
\end{tabular}
\end{table*}
\begin{table*} 
 \tiny
 \centering
\caption{The statistical results of the mean rank of The CCS/MOEAs and DCS/MOEAs. To simplify, the mean and variance values involve both IGD and IH of the corresponding three MOEAs.}
\label{Statistically Analysis}
\begin{tabular}{*{9}{c}}
\hline
\multirow{3}{*}{Algs.}&\multicolumn{4}{c}{First Five Instances}&\multicolumn{4}{c}{Last Fifteen Instances}\\
\cline{2-9}
&\multicolumn{2}{c}{1st Constraints}&\multicolumn{2}{c}{2nd Constraints}&\multicolumn{2}{c}{1st Constraints}&\multicolumn{2}{c}{2nd Constraints}\\
\cline{2-9}
&\multicolumn{1}{c}{Mean}&Var&\multicolumn{1}{c}{Mean}&Var&\multicolumn{1}{c}{Mean}&Var&\multicolumn{1}{c}{Mean}&\multicolumn{1}{c}{Var}\\\hline\hline
\multicolumn{1}{c}{CCS/MOEAs}&2.27&0.89&\multicolumn{1}{c}{2.60}&0.64&\multicolumn{1}{c}{3.00}&0.13&\multicolumn{1}{c}{3.10}&\multicolumn{1}{c}{0.28}\\\hline\multicolumn{1}{c}{DCS/MOEAs}&5.33&1.97&\multicolumn{1}{c}{4.80}&4.12&\multicolumn{1}{c}{5.17}&2.41&\multicolumn{1}{c}{4.37}&\multicolumn{1}{c}{3.45}\\\hline\hline

\end{tabular}
\end{table*}
\subsection{More Experiment Instances}
\label{section_7}

So far, we have only presented the results on OR-Libary with the \emph{first} constraint set. Furthermore, we provide more simulation experiments for this paper. We only analyze the statistical results here and the detailed tables are placed in the Appendix due to the space limit.  In Table~\ref{NGINX_IGD_Constraints1}, in terms of IGD, the performance of the CCS/MOEAs is not so outstanding as above. Although all of the CCS ones do not perform worse than the DCS ones, MODEwAwL performs similarly to CCS/MOEA/D and CCS/NSGA-\uppercase\expandafter{\romannumeral2} concerning both mean ranks and Wilcoxon rank-sum test. In Table~\ref{NGINX_IH_Constraints1}, the table about IH, the CCS/MOEAs still works better than DCS/MOEAs and MODEwAwL in most instances. It is worth noting that the CCS/MOEAs can not outperform MODEwAwL on all the problems since it is almost impossible for a heuristic algorithm to outperform all other heuristics. For example on D7 and D12, details presenting in the Appendix, MODEwAwL does the best in terms of IGD and IH.

Further, as for the global statistics, Table~\ref{Statistically Analysis} shows that the mean and variance values of the CCS/MOEAs are better than those of the DCS/MOEAs in all the cases. The lower mean values suggest that the CCS/MOEAs outperform the DCS/MOEAs on searching the optimal solutions of the constrained portfolio optimization. On the other hand, the lower variance values imply that the CCS/MOEAs are more robust. Besides, the statistical results of MOEAwAwL are not listed, because they are worse than all CCS/MOEAs in most cases.

Besides, we conduct some experiments with different population sizes, $NP=\ $250 and 500, since heavier runs of EAs may lead to better solutions~\cite{mavrotas2015improved}. The details are presented in the Appendix. There are two important results, (i) the CCS ones still outperform the DCS ones, and (ii) there is still room for improvement concerning IGD and IH.

To summarize, the comparison study approves CCS can provide better and robust solutions.

\section{Conclusions}
\label{section_6}
This paper studies a portfolio optimization problem, which can be modeled as an MINLP problem. In the literature, a variety of work has been done on how to deal with either multi-objective optimization problems or constrained optimization problems while mixed-variable optimization has not attracted much attention for EAs. It is not the first time using a real-valued vector to represent the mixed variables simultaneously in the constrained portfolio problems, but the difference between DCS and CCS is firstly discussed. It is pointed out that CCS is expected to utilize the dependence among variables during the search of EAs. Moreover, two tailored reproduction operators are proposed while the constrained portfolio optimization is a specific problem, which could be solved more efficiently with the prior information of the problem. Then the coding scheme, the search operators, and the constraint handling strategy are integrated into three major categories of multi-objective evolutionary algorithms. These new algorithms are conducted on 20 benchmark problems with different asset numbers. The comparison study among a state-of-the-art algorithm and the MOEAs has demonstrated that the new coding scheme, CCS, is promising for dealing with the constrained portfolio optimization.

In this paper, CCS is only applied to constrained portfolio optimization. There are still a variety of other problems with mixed variables worth exploring. Meanwhile, a more challenging problem, discussing why and how does the dependence among variables affect the search of EAs, is also expectant.

\bibliographystyle{elsarticle-num.bst}
\bibliography{Overall_Reference}
\end{document}